\documentclass[11pt, oneside]{article}  

\usepackage{geometry}  
\usepackage{float}
\usepackage{cmbright}
\usepackage{cite}
\usepackage{empheq} 

\usepackage{bookmark}
\usepackage{graphicx}							
\usepackage[breakable]{tcolorbox}
\usepackage[thinc]{esdiff} 
\usepackage{amssymb}
\usepackage{longtable}

\usepackage{amsmath}
\usepackage{booktabs}
\usepackage{color}
\definecolor{c1}{rgb}{0.12, 0.56, 1.0}
\usepackage{xspace}
\usepackage{bm}
\usepackage{xcolor}
\usepackage{caption}
\usepackage{subcaption}
\usepackage{authblk} 

\usepackage{sectsty} 
\definecolor{cool_blue}{RGB}{24, 132, 193}
\sectionfont{\color{c1}\selectfont}
\subsectionfont{\color{c1}\selectfont}
\subsubsectionfont{\color{c1}\selectfont}
\subparagraphfont{\color{gray}\selectfont}

\definecolor{fruitpushorange}{RGB}{255, 127, 0}

\newtcolorbox[auto counter
]{mybox}[2][]{
title=Box~\thetcbcounter: #2,#1,
}

\usepackage{hyperref} 

\newcommand{\hoft}{\ensuremath{h_{\bm \theta}(t; t_0, h_0)}\xspace}

\title{
Bayesian reversal of the liquid level trajectory in a draining tank for pollution forensics
}
\author[2]{Kyla D. Jones}
\author[1]{Gbenga Fabusola}
\author[2]{Alexander W. Dowling}
\author[1]{Cory M. Simon$^*$}
\affil[1]{School of Chemical, Biological, and Environmental Engineering. Oregon State University. Corvallis, OR. USA.}
\affil[2]{Department of Chemical and Biomolecular Engineering. University of Notre Dame. Notre Dame, IN. USA.}
\affil[$^*$]{\texttt{cory.simon@oregonstate.edu}}

\begin{document}
\maketitle

\begin{abstract}
Storage tanks for hazardous liquids are common in industry and agriculture. 
During a pollution incident, liquid may drain from a storage tank through a small hole, crack, or pipe. 
After containing the leak, estimating the discharged volume of liquid is essential for public safety, regulatory assessment, and remediation. 
When the original inventory of liquid is unknown, this constitutes an inverse problem. 
In this work, we present a framework for inferring the initial liquid level in a partially drained tank from the observed final liquid level after a pollution incident and an estimate of the drainage duration.
Because the drainage dynamics, model parameters, and observations are uncertain, we employ Bayesian statistical inversion to combine prior physical knowledge with experimental liquid level time series data to predict the initial liquid level with quantified uncertainty.
We use a physics-based model based on Torricelli's law to describe the tank-draining dynamics and augment it with an empirical discrepancy function to account for missing or imperfectly modeled physics. 
In our experiments with a tank draining of water, we found that our inferred initial liquid level was accurate, although uncertainty increased with drainage duration. 
Beyond its application to pollution forensics, this work may also serve as a hands-on classroom project illustrating dynamic modeling, model discrepancy, and Bayesian inference.
\end{abstract}

\clearpage


\section{Introduction}
Large above- or under-ground tanks are widely used in the chemical and petroleum industries to store hazardous liquids such as solvents, additives, precursors, intermediate or finished products, and waste~\cite{huang2013technical,pullarcot2015above}. 
In agriculture, above-ground tanks on farms store liquid fertilizers, fuel, pesticides, and livestock slurry~\cite{travnivcek2019prevention}. 
The wine industry ferments and stores wine in large tanks~\cite{montalvo2021environmental}.
Additionally, the U.S. Department of Energy's Hanford Site has more than 100 underground tanks storing radioactive and chemical waste generated during the separation of plutonium from irradiated fuel rods for nuclear weapons production~\cite{hanford}. 
At Fukushima~\cite{sylvester2013radioactive}, about 1000 tanks store cooling water, rainwater runoff, and groundwater contaminated with radioactive Tritium (as of 2022)~\cite{iea}. 

Equipment failure, operator error, negligence, and sabotage have caused unintended liquid discharges from storage tanks~\cite{chang2006study,hongguang2014fault,guerin2014understanding,travnivcek2019prevention}. 
These incidents can cause illness, injury, disability, or death among exposed personnel~\cite{bongers2008challenges,cullinan2002epidemiological}.
Further, soil, groundwater, and surface water can be contaminated, negatively affecting ecosystems, food, and drinking water~\cite{mclaughlin2016spills,doi:10.1021/acsomega.3c05187}.
Thus, regulatory, liability, and legal consequences can follow. 
Engineers can reduce the risk of accidental releases by designing tanks appropriately (e.g., materials selection) and performing routine maintenance and inspections~\cite{chang2006study}. 
Additionally, installing chemical sensors near storage tanks enables early leak detection and prompt responses such as discharge stoppage, spread containment, evacuation, and cleanup~\cite{ramirez1996detection,ho2001review,ly2021site}.

This paper concerns incident forensics for a storage tank that partially drained under gravity through a small hole, crack, or discharge pipe.
Estimating the volume of liquid discharged is important for assessing the degree of environmental contamination. 
When the initial liquid inventory is unknown, reconstructing the initial liquid level from the observed final liquid level and the estimated drain time becomes an inverse problem~\cite{groetsch1993inverse_tl}. Conceptually, we want to reverse the observed final liquid level backwards in time, to the estimated beginning of the drainage event.
Quantifying uncertainty in the inferred initial liquid inventory is also important because decisions regarding remediation, fines, and safety may depend on it. 

To address this problem, we employ a probabilistic framework that combines a forward model of the drainage dynamics with a stochastic observation model. 
Bayesian calibration methods~\cite{mebane2023bayesian,hou2021review,mebane2013bayesian,johnston2014updating} provide a natural approach for incorporating prior physical knowledge, experimental liquid level time series data, model discrepancy, and observation uncertainty within a unified inference framework. 
Bayesian statistical inversion and Markov chain Monte Carlo methods~\cite{kaipio2006statistical,calvetti2018inverse,dashti2015bayesian,waqar2023tutorial} then enable joint inference of uncertain model parameters, observation noise, and the unknown initial liquid level associated with the pollution event.

This work provides a practical framework for pollution forensics involving drainage incidents in storage tanks. 
More fundamentally, it contributes to the study of inverse problems involving draining tanks~\cite{groetsch1993inverse_tl,aberman2017dip,fabusola2025inferring}. 
Given the accessibility of the experiment and problem setting, the problem may also serve as a hands-on classroom project illustrating concepts in dynamic modeling, inverse problems, model discrepancy, and Bayesian inference~\cite{sidebotham2025workshops,martinez2022notion,powell2012carrying}.


\section{Problem Statement} \label{sec:inv_prob}

\begin{figure}[h!]
	\centering
	\includegraphics[width=0.7\textwidth]{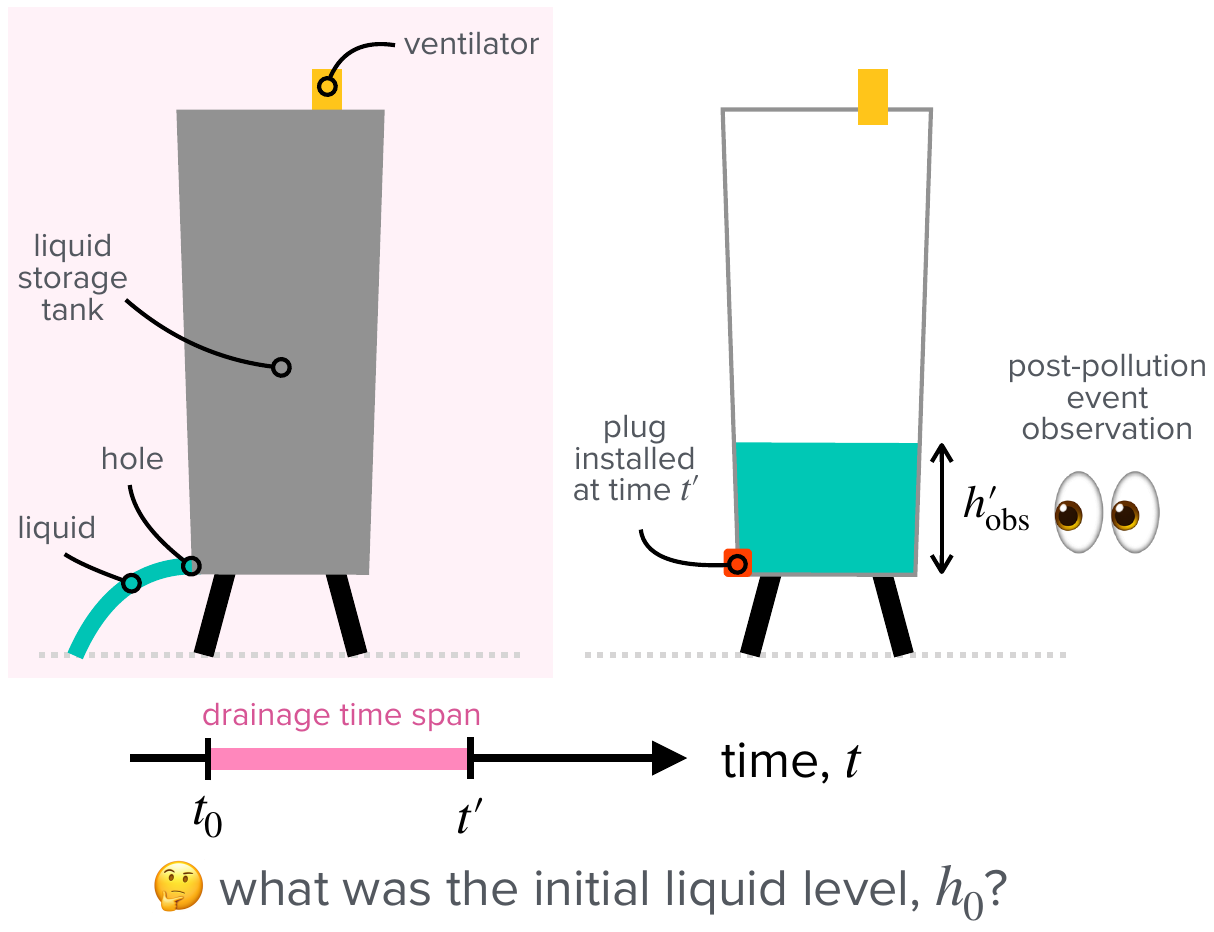}
	\caption{\textbf{Reconstructing a tank drainage event.} Liquid drains from a storage tank, under the force of gravity, through a small orifice (in the side of the tank and at the bottom) over the time interval $[t_0,t^\prime]$. The objective is to infer the unknown initial liquid level $h_0$ from the observed final liquid level $h^\prime_{\rm obs}$ and the drainage duration.}\label{fig:setup}
\end{figure}

Fig.~\ref{fig:setup} summarizes the problem setup.
An above-ground storage tank holds a hazardous liquid.
The initial height of the liquid in the tank, $h_0$~[cm], is unknown.
At time $t=t_0$~[s], the tank suddenly begins draining through a small orifice near its bottom due to the hydrostatic pressure generated by the liquid column caused by gravity, contaminating the surrounding area.
Owing to a vent, the headspace in the tank remains at atmospheric pressure during draining.
An \textit{in situ} gas sensor outside the tank alerts onsite personnel to the discharge.
At time $t^\prime > t_0$, the personnel seal the orifice to stop the drainage.
The remaining height of liquid, $h^\prime_{\rm obs}$~[cm], is measured.
Inspection of sensor logs and video surveillance gives an estimate of the drainage time span $[t_0, t^\prime]$.
The objective is to estimate, with quantified uncertainty, the initial height of liquid $h_0$ in the tank to calculate the volume of hazardous liquid released into the environment.

\section{Methods}

\subsection{Experimental setup and water level time series data collection}

The tank used in the experiments was an open-top plastic kitchen container approximately shaped as a truncated inverted square pyramid. 
To allow drainage, a small hole was drilled into the side of the tank near its base. 
To monitor the liquid level over time, centimeter-scale markings were placed along the tank's side.

Each experiment began by filling the tank with tap water to a prescribed initial liquid level, with the drainage hole plugged. 
At time $t = 0$, the plug was removed to initiate gravity-driven drainage through the orifice. 
The liquid level was then recorded over time as the tank drained.

\paragraph{Liquid level time series data.} 
We conduct and generate liquid level time series data from three replicate tank-draining experiments.

The first experiment aims to mimic the pollution scenario in Fig.~\ref{fig:setup} by generating a single data point concerning the final liquid level:
\begin{equation*}
	\mathcal{D}^\prime = \{ (t^\prime, h_{\rm obs}(t^\prime))\}.
\end{equation*} 
We note the initial condition $(0, h_{0, \rm obs})$, but hold it out from our inference procedure to test our ability to accurately solve the inverse problem.

The purpose of the second and third experiments is to calibrate the forward model of the liquid level dynamics in the tank. 
In these experiments, we fill the tank to the top, then record the liquid level over time to give two time series data sets that trace the trajectory of the liquid level in the tank over the entire process of draining:
\begin{equation*}
	\mathcal{D}^{(k)} = \{ (t_i^{(k)}, h_{\rm obs}(t_i^{(k)}) ) \}_{i=0}^{N},\quad k \in \{1,2\}.
\end{equation*}
We omit data at the smallest liquid levels because surface tension stopped outflow.
We pool the liquid level time series data from all three experiments into one data set
\begin{equation*}
	\mathcal{D} = \mathcal{D}^\prime \cup \mathcal{D}^{(1)} \cup \mathcal{D}^{(2)}
\end{equation*}
to jointly infer the forward model parameters and the initial condition in the pollution scenario.

\subsection{Data-generating model} 
\label{sec:kohmethods}
Our data-generating model for the observed liquid level dynamics during tank drainage combines a physics-based dynamic model, an empirical model discrepancy function, and a probabilistic observation-noise model. 
This formulation is commonplace in Bayesian calibration and statistical inversion for computer models~\cite{kennedy2001bayesian,brynjarsdottir2014learning,calvetti2018inverse,schenk2026framework}. 
Specifically, we model the observed liquid level $H_{\rm obs}(t)$ at time $t \in [t_0,t^\prime]$ as a random variable:
\begin{equation}
        H_{\rm obs}(t) \mid t_0, h_0, \bm \theta, \bm a , \sigma = h_{\bm 	\theta }(t; t_0, h_0) + \delta_{\bm a} ( h_{\bm \theta }(t; t_0, h_0) ) + E_\sigma. \label{eq:forward}
\end{equation}

The first term in eqn.~\eqref{eq:forward}, $h_{\bm \theta }(t; t_0, h_0)$, is a physics-based model for the liquid level at time $t$. 
Evaluating this model requires both the initial condition $(t_0,h_0)$ and the vector of physically meaningful model parameters $\bm \theta$.

The second term, $\delta_{\bm a}(h_{\bm \theta}(t; t_0, h_0))$, is an empirical discrepancy function that accounts for systematic differences between the physics-based model and the true liquid level. 
This term captures the effects of unmodeled physics, idealized assumptions, and structural inadequacies in $h_{\bm \theta}(t; t_0, h_0)$.
The vector $\bm a$ parameterizes the discrepancy function, although its coefficients are not necessarily physically interpretable. 
Consequently, the sum $h_{\bm \theta }(t; t_0, h_0) + \delta_{\bm a}(h_{\bm \theta}(t; t_0, h_0))$ represents the true, but unobserved, liquid level at time $t$.

The final term, $E_\sigma$, represents measurement noise contaminating the observed liquid level. 
We assume the measurement noise is independent and identically distributed across observations, with mean zero and unknown variance $\sigma^2$.

The following sections describe each component of the data-generating model in detail.

\subsubsection{Physics-based dynamic model of the liquid level in the draining tank}
Now, we specify the physics-based model $h_{\bm \theta }(t; t_0, h_0)$ of the dynamics of the liquid level in the tank.

The dynamics of the liquid level in the tank depend on the tank geometry, which we approximate as a truncated inverted square pyramid. 
We assume that the liquid surface remains flat during the draining process. 
The tank has height $h_{\max}$~[cm], top side length $x_t$~[cm], and bottom side length $x_b$~[cm].
Thus, the horizontal cross-sectional area of the tank at height $h \in [0, h_{\rm max}]$, which dictates the dynamics of draining, is given by
\begin{gather*}
        \alpha (h) = \left[\frac{h}{h_{\max}} x_t + \left(1-\frac{h}{h_{\max}} \right)x_b\right]^2.
\end{gather*}

The outflow rate depends on the orifice geometry.
The orifice is circular, with radius $r$~[cm], and located at the bottom side of the tank. 
We model the outflow velocity as a function of liquid level $h$ as $c\sqrt{2gh}$ according to Torricelli's law, which treats the liquid as incompressible and inviscid and the orifice as negligibly small relative to the tank's girth.
Here, $g$~[cm/s$^2$] is the acceleration due to gravity and the unitless $c \in [0,1]$ is a discharge coefficient.

Application of a mass balance of [incompressible] liquid in the tank, incorporating Torricelli's law and the tank geometry, yields the dynamic model for the liquid level in the tank~\cite{debook,fabusola2025inferring,rother2024modelling,liu2008drainage,belinskiy2020time,powell2012carrying,Calero2008,pavesi2019investigating,williams2021vessel}:
\begin{subequations} \label{eq:ivp}
        \begin{empheq}[left=\empheqlbrace]{align}
                &\alpha (h(t)) \diff{h(t)}{t} =  - c\pi r^2 \sqrt{2gh(t) }, \quad t \in [t_0, t^\prime], \label{eq:forward_model} \\
                &h(t_0) = h_0.
                \label{eq:ic}
        \end{empheq}
\end{subequations}
The term on the left side of the ordinary differential equation is the rate of change of the volume of liquid in the tank, while the term on the right is the volumetric outflow rate.
We solve this nonlinear initial value problem numerically.
Its solution is the physics-based model for liquid level trajectory $h_{\bm \theta}(t; t_0, h_0)$, where $\bm \theta = (h_{\max}, x_t, x_b, c, r)$ lists the model parameters while $g$ is assumed to be known.

\subsubsection{Model discrepancy} 
The functional form of the empirical discrepancy function $\delta_{\bm a}(h)$ is generally unknown and depends on the particular modeling problem. 
To flexibly represent the discrepancy, we express $\delta_{\bm a}(h)$ using degree-$n$ Bernstein basis polynomials~\cite{lorentz2012bernstein}:
\begin{equation*}
        \delta_{\bm a} ( h ) = \sum_{\nu=0}^{n} a_{\nu} \binom{n}{\nu }
        \left(\frac{h}{h_{\rm max}}\right)^\nu
        \left(1-\frac{h}{h_{\rm max}}\right)^{n-\nu}.
\end{equation*}
Thus, the discrepancy function is parameterized by the coefficient vector $\bm a \in \mathbb{R}^{n+1}$.

Bernstein basis polynomials provide a flexible and interpretable representation of smooth discrepancy functions over a bounded domain.
Each basis function has domain and range $[0,1]$.
Fig.~\ref{fig:bern} displays the degree-$2$ Bernstein basis polynomials used in this work.
The coefficient $a_\nu$ can be interpreted loosely as controlling the discrepancy near the relative liquid level where the $\nu$-th Bernstein polynomial attains its maximum.
For example, when $n=2$, the coefficient $a_1$ primarily influences the discrepancy near $h/h_{\rm max}=0.5$, corresponding to a half-full tank.
Likewise, the sign of $a_\nu$ is somewhat indicative of whether the physics-based model underpredicts or overpredicts the true liquid level in that region.

Although the coefficients $\bm a$ admit some qualitative interpretation, the modeler can also interpret the discrepancy function by visualizing of $\delta_{\bm a}(h)$ itself.
Moreover, each Bernstein basis polynomial has support over the entire domain, so individual coefficients influence the discrepancy function globally rather than locally.

Importantly, we express the discrepancy as a function of the predicted liquid level rather than as a function of time.
This choice reflects the fact that the drainage process is time-invariant, since the governing dynamics for $h(t)$ do not explicitly depend on time.

\subsubsection{Observation/measurement noise}
The observation/measurement noise is assumed to be Gaussian $E_{\sigma} \mid \sigma \sim \mathcal{N}(0, \sigma^2)$.

\subsubsection{The likelihood density}
The data-generating model combines the physics-based dynamic model of the liquid level in the tank, the empirical model discrepancy function, and observation/measurement noise and is summarized by the conditional probability distribution corresponding with eqn.~\ref{eq:forward}:
\begin{equation}
\begin{split}
p(h_{\rm obs}(t) \mid t_0, h_0, \bm \theta, \bm a, \sigma) = \\
    \frac{1}{\sqrt{2\pi}\sigma} 
    \exp \left[
        -\frac{1}{2}
        \left(
	\frac{
            \hoft + \delta_{\bm a}(\hoft) - h_{\rm obs}(t)
        }{
            \sigma
        }
	\right)^2
    \right].
\end{split}
\label{eq:likelihood_point}
\end{equation}
The forward model being probabilistic captures the aleatory uncertainty arising from irreducible randomness over the process of draining and our observations of the liquid level.

\subsection{Bayesian inference of the forward model parameters and initial condition}
Adopting a Bayesian perspective, we treat the physics-based model parameters, discrepancy function parameters, measurement noise variance, and initial conditions for the pollution event and the two draining experiments for calibration 
\begin{equation*}
	\bm \beta = (\bm \theta, \bm a, \sigma, t_0, h_0, (t_0^{(k)}, h_0^{(k)})_{k\in\{1,2\}})
\end{equation*}
as random variables.
The probability density of $\bm \beta$ captures epistemic uncertainty about these parameters and initial conditions arising from limited data.
It expresses our degree of belief about the values of these quantities based on available information and data.

\paragraph{Prior density.}
First, we specify the joint prior density $p(\bm \beta)$ to encode information and beliefs about the parameters and initial conditions that are available before observing any liquid level time series data over the process of the tank draining.
The prior for the model parameters is informed by measurements of the tank geometry and typical values of the discharge coefficient. 
The prior for the model discrepancy parameters reflects our judgement of the accuracy of the physics-based model, while the prior for the measurement-noise variance reflects the precision of our visual observations of the liquid level. 
For the initial conditions, the pollution scenario prior is based on our estimate of the drainage start time and our belief about plausible initial liquid levels. 
For the calibration experiments, the priors reflect the controlled initial conditions used in the experiments.

Generally, a prior can range from informative, to weakly informative, to diffuse based on the concentration of its density.
A diffuse prior, such as a uniform distribution over a large range, represents a high degree of uncertainty about the value of a variable. 
By contrast, an informative prior, like a sharply peaked Gaussian distribution, indicates a high degree of certainty about the value of a variable.~\cite{van2021bayesian}
Our priors over the dimensions of the tank and initial liquid levels in the calibration experiments are informative because they are based on direct (but noisy) measurements. 
Meanwhile, our prior for the initial liquid level for the pollution scenario is diffuse to reflect high uncertainty about this unknown. 
A diffuse prior allows the liquid level data to more strongly inform the posterior distribution.
If instead, we possessed logs of past liquid inventories, we could construct a more informative prior for the initial liquid level.

\paragraph{Likelihood function.}
The likelihood $p(\mathcal{D} \mid \bm \beta)$ is the probability density of the liquid level time series data $\mathcal{D}$, conditioned on a particular value for the forward model parameters and initial conditions in $\bm \beta$.
Under the probabilistic forward model in eqn.~\ref{eq:likelihood_point}, the likelihood is:
\begin{equation*}
	p(\mathcal{D} \mid \bm \beta ) =
		p(h_{{\rm obs}}(t^\prime ) \mid t_0, h_0, \bm \theta, \bm a,  \sigma )
		 \prod_{k \in \{1,2\}} \prod_{i=0}^N p(h_{{\rm obs}}(t_i^{(k)}) \mid t_0^{(k)}, h_0^{(k)}, \bm \theta, \bm a,  \sigma).
\end{equation*}
Viewed as a function of the parameter and initial condition vector $\bm \beta$ after the data $\mathcal{D}$ are collected, the likelihood scores each $\bm \beta$ according to its consistency with the data $\mathcal{D}$ under the data-generating model with that $\bm \beta$. 
The first factor describes consistency with the final condition of the liquid level observed in the pollution scenario.
The second two factors describe consistency with the liquid level time series data collected from the two calibration experiments.

\paragraph{Posterior.}
Bayes' theorem combines the likelihood and the prior to yield the posterior distribution of the parameters and initial conditions~\cite{van2021bayesian}
\begin{equation*}
	p(\bm \beta \mid \mathcal{D}) = \frac{p(\mathcal{D} \mid \bm \beta )\, p(\bm \beta)}{p(\mathcal{D})},
\end{equation*}
where
\begin{equation}
	p(\mathcal{D}) = \int p(\mathcal{D} \mid \bm \beta )\, p(\bm \beta )\, d\bm \beta \label{eq:evidence}
\end{equation}
is the marginal likelihood or evidence.
The posterior distribution of the forward model parameters and initial conditions represents our updated beliefs about them, after comparing the liquid level time series data $\mathcal{D}$ with the forward model.

\paragraph{Sampling from the posterior.}
The evidence integral (eqn.~\ref{eq:evidence}) is generally analytically intractable, and closed-form expressions for the posterior are unavailable unless conjugate priors are selected~\cite{liu2014bayes}.
Consequently, direct evaluation of the posterior density is generally not possible.
Instead, we resort to Markov chain Monte Carlo (MCMC) sampling to approximate the posterior.

MCMC methods generate a realization of a Markov chain $( \bm \beta^{(\ell)})_{\ell=0}^L$ whose stationary distribution matches the posterior distribution $p(\bm \beta \mid \mathcal{D})$.
Sampling a Markov chain relies on ratios of posterior densities, 
\begin{equation*}
	\frac{p(\bm \beta^{(\ell+1)} \mid \mathcal{D})}{p(\bm \beta^{(\ell)} \mid \mathcal{D})} = \frac{p(\mathcal{D} \mid \bm \beta^{(\ell+1)}) p(\bm \beta^{(\ell+1)})}{p(\mathcal{D} \mid \bm \beta^{(\ell)}) p(\bm \beta^{(\ell)})},
\end{equation*}
so the evidence term $p(\mathcal{D})$ need not be computed. 

We use a Hamiltonian~\cite{betancourt2017conceptual} MCMC sampler~\cite{van2021bayesian}, particularly the No-U-Turn Sampler (NUTS)~\cite{hoffman2014no}.
Four independent chains were run in parallel for 5{,}000 iterations each. 
The first third of the chains were discarded as burn-in samples. 
We assessed convergence quantitatively and qualitatively. 
Quantitatively, we computed the Gelman–Rubin and Brooks diagnostics~\cite{gelman1992inference,brooks1998general} which compare within- and between-chain variances of each parameter.~\cite{roy2020convergence}   
Qualitatively, we evaluated trace plots to assess thorough exploration of the credible interval and histograms to assess agreement between posteriors from each Markov chain.

\subsection{Software}
We conducted our modeling and inference in the Julia programming language~\cite{bezanson2017julia}.
We used a numerical solver in \texttt{DifferentialEquations.jl}~\cite{rackauckas2017differentialequations} for the initial value problem and the probabilistic programming package \texttt{Turing.jl}~\cite{ge2018turing} for MCMC sampling.


\section{Results}
We now tackle an instance of the inverse problem of time reversal described in Sec.~\ref{sec:inv_prob} through experiments with an open-top tank draining of water, under the force of gravity, through a small orifice in its side at the bottom.
Tap water serves as a safe surrogate for contaminated water and various inviscid liquid pollutants.

\subsection{Problem instance}
\begin{figure}[!htbp]
        \centering
          \begin{subfigure}[b]{0.375\textwidth}
        		\includegraphics[width=\textwidth]{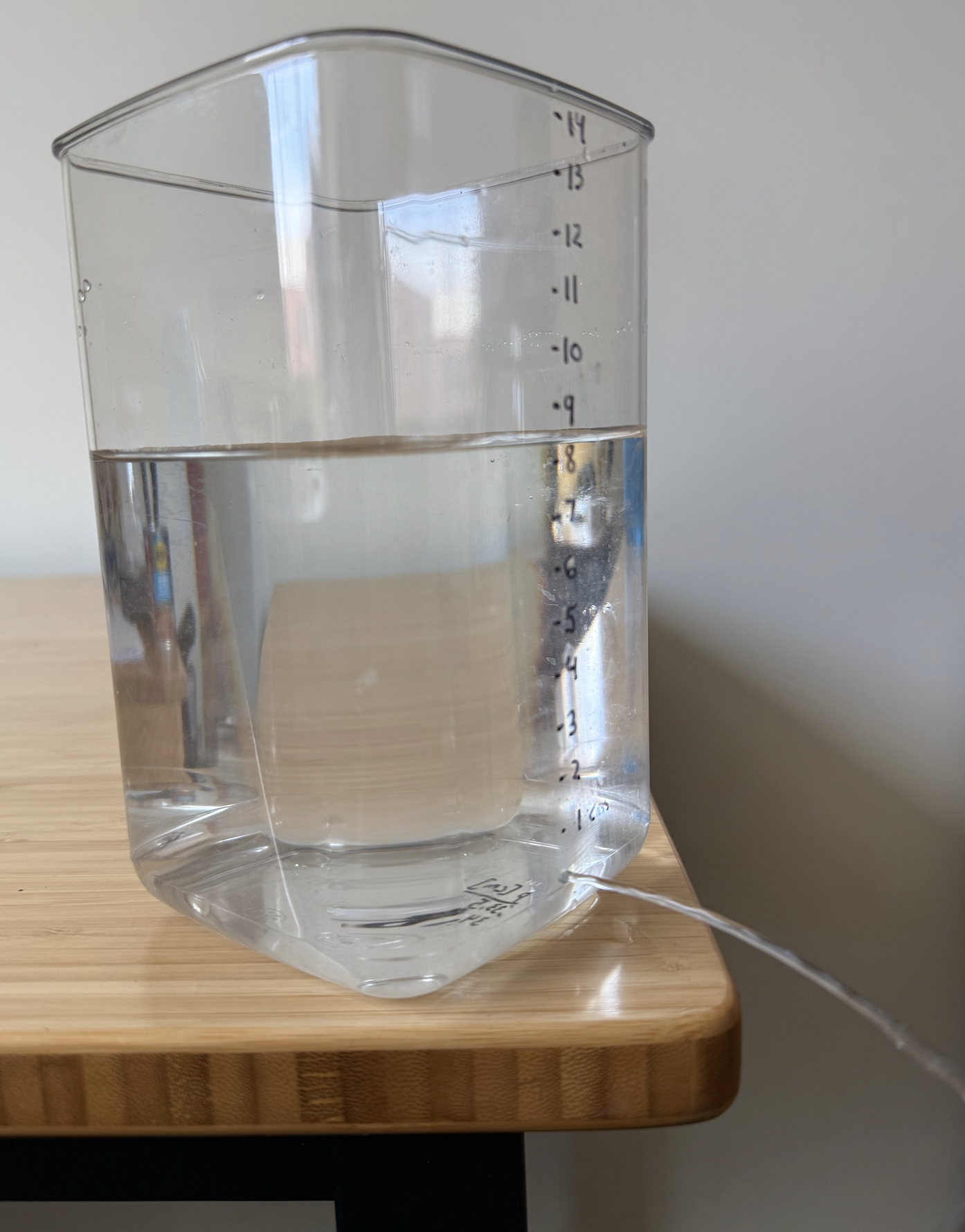}
        		\caption{Experimental setup} \label{fig:expsetup}
        \end{subfigure}
        
        \begin{subfigure}[b]{0.825\textwidth}
        		\includegraphics[width=\textwidth]{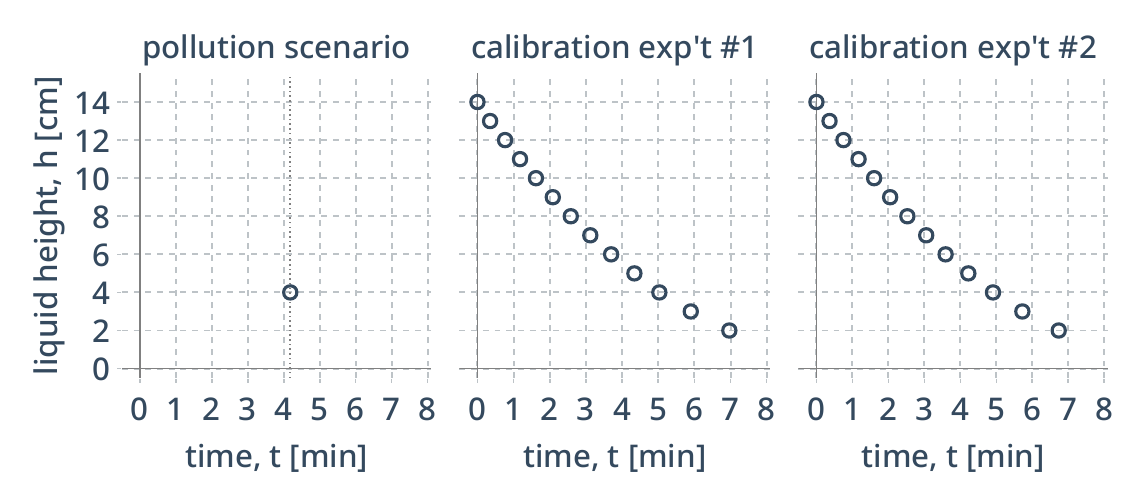}
        		\caption{Liquid level time series data} \label{fig:data}
        \end{subfigure}
        \caption{\textbf{Experimental setup and liquid level time series data.}
       (a) Water drains out of a small hole in the bottom side of a tank. 
       (b) The measured height of water in the tank over time during three separate tank-draining experiments. The first experiment mimics a pollution event, where only the final condition is observed. The next two experiments are for model calibration. Beginning with a filled tank, the liquid level time series data trace the entire trajectory of the liquid level over the process of draining.
       }
\end{figure}

Fig.~\ref{fig:expsetup} displays our experimental setup. 
Water drains through a small hole in a tank while we visually inspect the liquid level using markings on the tank and record time with a stopwatch.

Fig.~\ref{fig:data} displays the resulting water level time series data in the draining tank over the course of three separate experiments. 
The first experiment mimics a pollution event. 
We allowed water to drain from the tank through the orifice for an estimated 4.18\,min, with $\pm$5\,s uncertainty.
After the hole was plugged to stop the drainage, we measured the remaining water level in the tank as 4\,cm.
Thus, the post-drainage event data point, representing the final condition, is $\mathcal{D}^\prime = (4.18~\text{min}, 4~\text{cm})$, with $t_0\approx 0$~min marking the beginning of the draining process.
For validation, we also measured the initial condition as $(0~\text{min}, 12~\text{cm})$, but conducted inference as if this data were unavailable.
The second and third experiments began with a full tank and generate time series data $\mathcal{D}^{(k)}$ for $k\in\{1,2\}$ of the water level to trace the water level trajectory over the full process of draining.
We conducted these experiments for the purpose of gathering data to calibrate the forward model of the water level dynamics in the tank.

\subsection{Bayesian inference}
Next, we leverage all water level time series data in Fig.~\ref{fig:data} to jointly calibrate the forward model and infer the initial condition during the pollution event.
Following the Bayesian approach, we first specify prior distributions of the parameters of the forward model and initial conditions in the three experiments, then use the water level time series data to update the prior to give a posterior distribution over the forward model parameters and initial conditions.
Ultimately, the solution to the inverse problem is given by the posterior distribution of the initial water level in the tank in the pollution scenario. 

\subsubsection{Prior distributions}
First, we specify our prior distributions for the model parameters, discrepancy parameters, observation/measurement noise variance, and initial conditions over the three tank drainage experiments.
The prior distributions are specified in detail in Sec.~\ref{sec:detailed_priors}. 
We briefly summarize them below.

\begin{figure}[!htbp]
        \centering
          \begin{subfigure}[b]{0.825\textwidth}
        		\includegraphics[width=\textwidth]{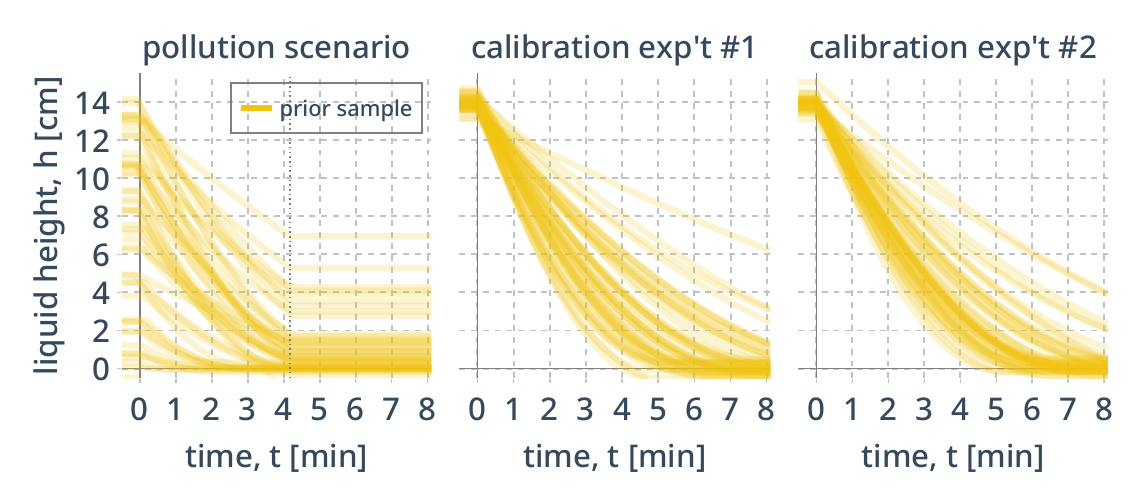}
        		\caption{Prior distribution.} \label{fig:prior}
        \end{subfigure}
        
        \begin{subfigure}[b]{0.825\textwidth}
        		\includegraphics[width=\textwidth]{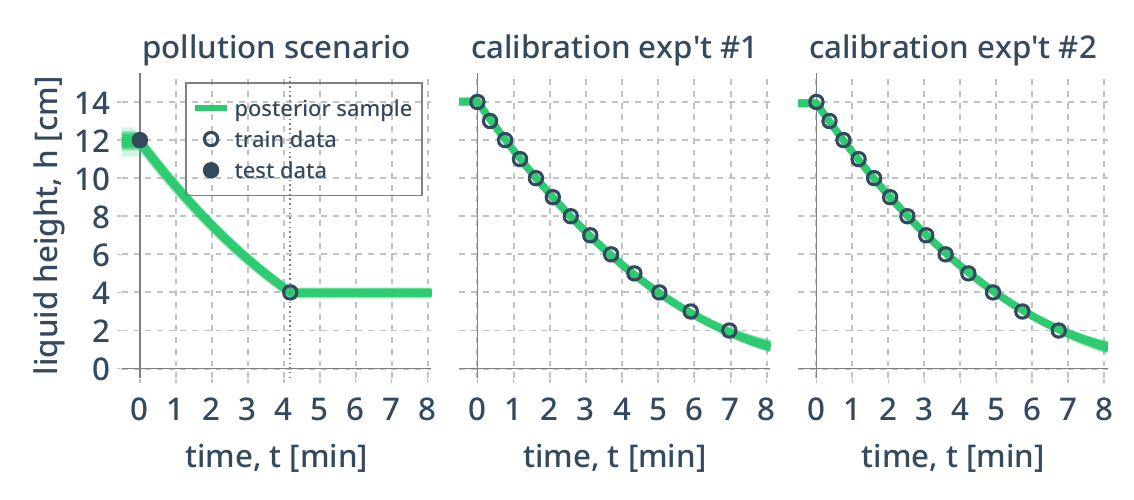}
        		\caption{Posterior distribution.} \label{fig:posterior}
        \end{subfigure}
        \caption{\textbf{The prior and posterior distribution over water level trajectories in the tank over the three experiments.} Each curve shows the forward model for the liquid level dynamics in the draining tank, $h_{\bm \theta }(t; t_0, h_0)+\delta_{\bm a}(h_{\bm \theta }(t; t_0, h_0))$, with parameters $\bm \theta$ and $\bm a$ and associated initial condition $(t_0, h_0)$ sampled from the (a) prior and (b) posterior. The liquid level time series data for obtaining the posterior are the hollow points in (b). The solid point represents the held-out initial condition for the pollution scenario.
       }
\end{figure}

The prior distribution over model parameters and initial conditions is summarized by samples of the corresponding predicted water level trajectories over the three experiments in Fig.~\ref{fig:prior}. 
For the pollution scenario, the backward water level trajectory is highly uncertain.
This uncertainty arises from both uncertainty about the model and discrepancy parameters and the initial water level.
For the two calibration experiments, which begin with a full tank, the water level trajectories reflect uncertainty largely about only the model and discrepancy parameters since the initial condition is set by us.

The marginal prior distribution over each model parameter is visualized in Fig.~\ref{fig:prior_posterior_params}.
Note, the priors over the dimensions of the tank and the orifice radius are informed by direct measurements and, hence, are quite concentrated.
Meanwhile, the weakly informative priors over the discharge coefficient, model discrepancy parameters, and measurement noise standard deviation are less concentrated. 

\paragraph{Tank geometry and orifice.} 
We specified prior distributions for the parameters describing the tank geometry---$x_b$, $x_t$, and $h_{\max}$---based on measurements of the dimensions of the tank with a length-measuring tape.
The prior for the radius of the orifice $r$ was informed by the radius of the drill bit used to create the hole. 
Specifically, we assigned $x_b$, $x_t$, $h_{\max}$, and $r$ Gaussian priors centered at their measured value with a small variance to reflect measurement uncertainty.

\paragraph{Discharge coefficient.}
We assigned the discharge coefficient $c$  a weakly informative prior distribution based on the range of discharge coefficients observed in previous studies. 
To enforce the physical constraint $c \in [0,1]$, we used a Beta prior distribution.

\paragraph{Measurement/observation noise.}
We adopt a weakly informative exponential prior on the standard deviation of the measurement noise $\sigma$, which is strictly positive, to reflect our assessment of the precision of the liquid level measurements.
We assign the prior for $\sigma$ a wide variance so that its value is primarily informed by the water level time series data.
\paragraph{Model discrepancy.} 
For the model discrepancy parameters $\bm a$, we impose ``spiky'' Laplace priors. 
Each prior is centered at zero to encourage sparsity and express our belief that the physics-based dynamic model of the water level during the draining process is accurate until the water level time series data inform us otherwise.
Additionally, we selected each variance to reflect the degree of model bias we think we might observe. 

\paragraph{Initial conditions.}
For the initial conditions of the three tank-draining experiments, we treat the pollution scenario distinctly from the calibration experiments.
In the pollution scenario, we use the rough estimate of the drainage duration to impose an informative Gaussian distribution for the drain start time $t_0$, centered at zero and with standard deviation of $5$\,s to reflect uncertainty.
To reflect no prior information about the initial water level, we assign a uniform prior distribution to $h_0$ over the interval from zero to the tank height, allowing the observation of the final liquid level to primarily inform inference of the initial water level.
Because the priors are specified before incorporating the observed water level data, the prior distribution for the initial liquid level assigns density to values below the observed final liquid level. 
After conditioning on the observations through the likelihood function, the posterior density is small on such values.
For the two calibration experiments, since we controlled the initial condition, we imposed informative Gaussian distributions on the drain start times $t_0^{(k)}$ and initial water levels $h_0^{(k)}$ for $k\in\{1,2\}$ at the values we set them in the experiment.

\subsubsection{Posterior distributions}
Next, we update the prior distributions for the forward model parameters, the measurement noise variance, and the unknown initial conditions using the water level time series shown in Fig.~\ref{fig:data}. 
The data comprise the observed final condition from the pollution event, $\mathcal{D}^\prime=\{(4.18~\text{min}, 4~\text{cm})\}$, and the two complete water level time series data sets $\mathcal{D}^{(k)}$ for $k\in\{1,2\}$ collected during controlled calibration experiments.

We jointly calibrate the forward model and infer the unknown initial condition associated with the pollution scenario.
To sample from the posterior distribution over the forward model parameters, measurement noise variance, and initial conditions across the three experiments, we employ the NUTS.
This posterior distribution of the model and discrepancy parameters, together with the model structure in eqn.~\ref{eq:forward}, constitute the calibrated data-generating model. 

Figs.~\ref{fig:ess},~\ref{fig:rhat}, and~\ref{fig:convergence} summarize the convergence diagnostics for the MCMC sampling. 
The Gelman--Rubin and Brooks statistics are consistent with convergence across chains, while the effective sample sizes indicate adequate posterior exploration. 
Additionally, the trace plots and marginal posterior histograms exhibit stable mixing and consistent stationary behavior across chains, supporting convergence to the posterior distribution.

\begin{figure}[!htbp]
        \centering
          \begin{subfigure}[b]{0.8\textwidth}
        		\includegraphics[width=\textwidth]{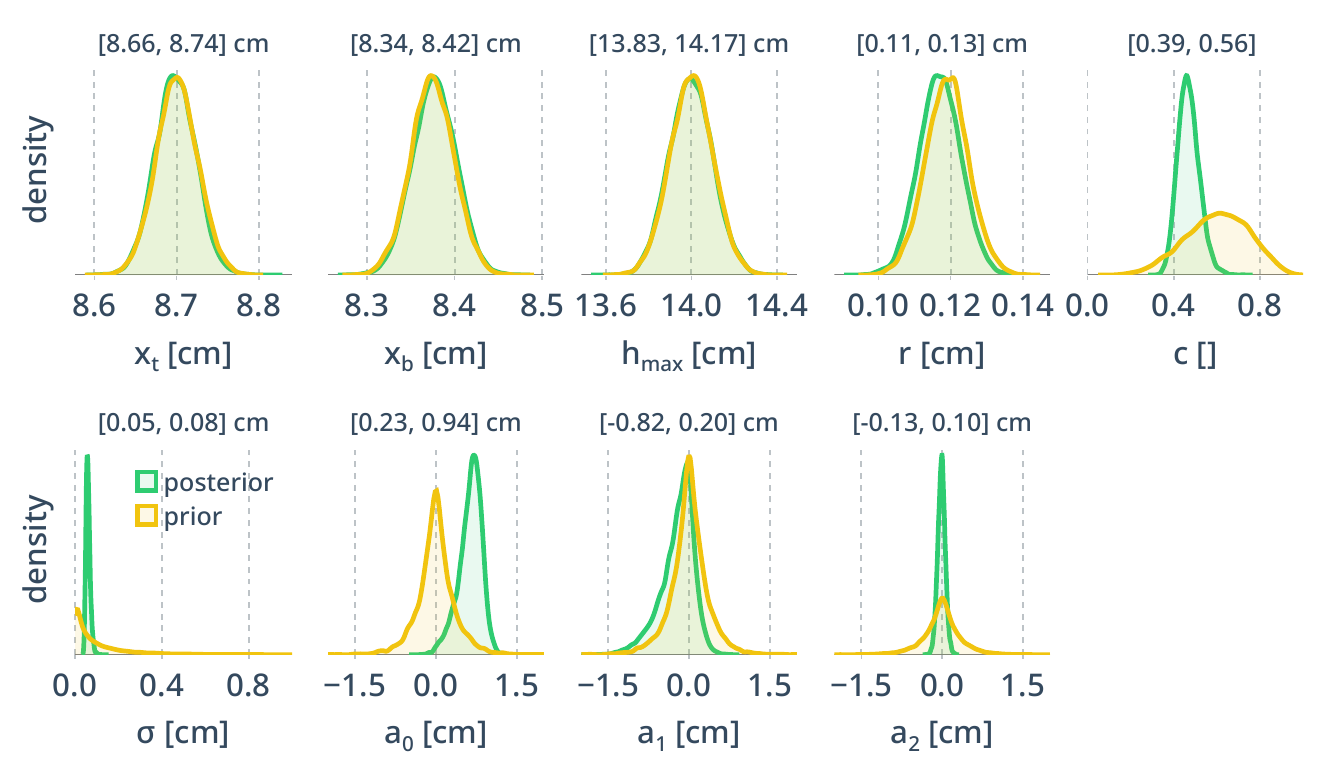}
        		\caption{The prior and posterior distribution of model parameters.} \label{fig:prior_posterior_params}
        \end{subfigure}
        
        \begin{subfigure}[b]{0.4\textwidth}
        		\includegraphics[width=\textwidth]{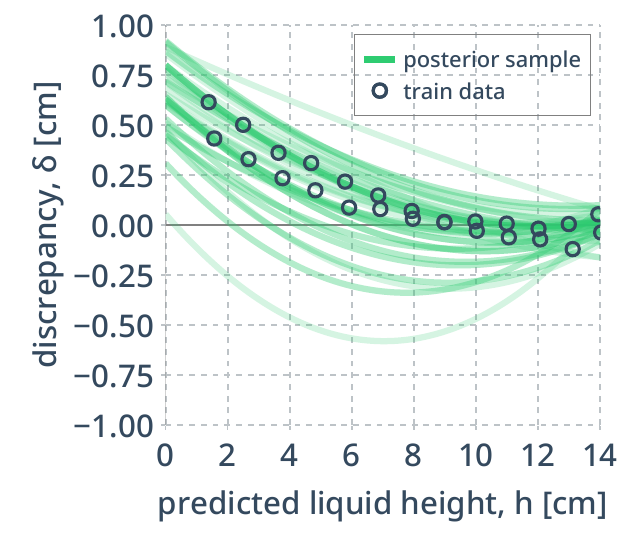}
        		\caption{The posterior model discrepancy.} \label{fig:model_disc}
        \end{subfigure}
        \caption{\textbf{Calibrated parameters of the forward model.} (a) Marginal prior and posterior densities, approximated by kernel density estimation on the MCMC samples, of each parameter of the forward model. Titles show equal-tailed 90\% credible interval. (b) Samples of the model discrepancy function from the posterior distribution, compared with average residuals for the water level time series data in the calibration experiments. 
       } \label{fig:calibration}
\end{figure}

\paragraph{The calibrated parameters of the forward model.}
Fig.~\ref{fig:prior_posterior_params} displays the marginal posterior densities of the data-generating model parameters, along with their prior densities. The posterior distributions are unimodal and that the credible intervals are approximately symmetric about their modes. Comparing the prior and posterior distributions in Fig.~\ref{fig:prior_posterior_params} illustrates the extent to which the observed data update the uncertain model parameters through Bayesian calibration. 
The posterior densities of the discharge coefficient, measurement noise standard deviation, and two of the model discrepancy parameters are more concentrated than their prior and/or shifted with respect to the corresponding prior densities. 
These changes reflect the information that the water level time series data provided about them. 
Meanwhile, the data did not provide additional information or contradict the priors placed on the tank dimensions. 

Fig.~\ref{fig:pairplot_calibration} shows joint marginal pair plots to visualize correlations between parameters in the posterior distribution. 
The first two coefficients of the discrepancy function are positively correlated, while the discharge coefficient and orifice radius exhibit a strong negative correlation.
This negative correlation makes sense because $c$ and $r$ are not independently identifiable from the product $cr^2$ appearing in eqn.~\ref{eq:forward_model}.

Fig.~\ref{fig:model_disc} shows the posterior discrepancy function obtained by sampling the posterior distributions of the quadratic Bernstein basis polynomial coefficients in Fig.~\ref{fig:prior_posterior_params}.
The inferred discrepancy function is consistent with the residuals between the calibration data and the physics-based model's posterior mean.
The discrepancy remains small while the tank drains from full to approximately half-empty. 
Near depletion, however, the discrepancy becomes positive, indicating that the physics-based model underpredicts the liquid level. 
In this regime, the discrepancy is at most approximately 1~\,cm. 
This behavior is consistent with experimentally observed slowing and eventual cessation of outflow near depletion, likely due to surface tension and viscous effects omitted from the physics-based model.

Finally, Fig.~\ref{fig:posterior} compares the posterior distribution of water level trajectories predicted by the forward model with the water level time series data. The calibrated forward model achieves strong predictive agreement with the observed water level time series data in each experiment.
This result is evidenced by the posterior predictive liquid level trajectories closely following the observations.
The mean absolute error between the data and posterior samples for the first and second calibration experiments is 0.20~\,cm and 0.11~\,cm, respectively.
For the pollution scenario, the water level trajectories pass nearby the observed final water level.
Taken as a whole, Fig.~\ref{fig:posterior} confirms that the calibrated model describes the dynamics of the draining process quite well, thereby reinforcing confidence in the predicted initial water level in the pollution reconstruction scenario.

\begin{figure}[!htbp]
        \centering
        	\includegraphics[width=0.8\textwidth]{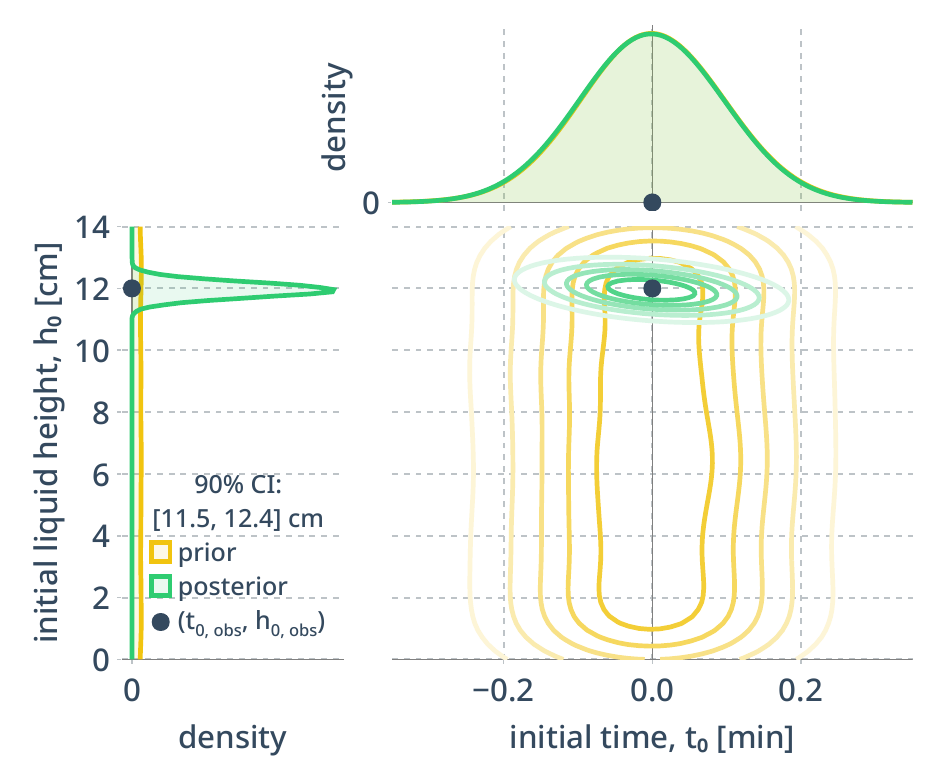}
        \caption{\textbf{The inferred initial condition of the draining tank for the pollution scenario.} 
        The joint (bottom right) and marginal (top right and bottom left) prior and posterior distributions of the initial condition of the water level in the tank during the pollution scenario. The point represents the true, held-out initial condition.
       }\label{fig:sol_inv_prob}
\end{figure}

\paragraph{Reconstructing the pollution scenario.}
The backward posterior water level trajectories in Fig.~\ref{fig:posterior} (pollution scenario) and the joint posterior distribution over the initial condition in the pollution scenario in Fig.~\ref{fig:sol_inv_prob} constitute the solution to the time reversal problem. Both show that the solution to the pollution scenario reconstruction problem is accurate and quantifies uncertainty. 

The backward liquid level trajectories Fig.~\ref{fig:posterior} (pollution scenario), conceptually, reverse the water level backward in time from the final observed water level to the predicted initial water level. We obtain each by simulating a forward model under model and discrepancy parameters sampled from the posterior distribution. The backwards trajectories cover the true held-out initial water level. 

Fig.~\ref{fig:sol_inv_prob} directly compares the prior and posterior distributions of the initial condition of the draining tank during the pollution scenario. The final liquid level concentrates around the true, held-out initial water level. 
The 90\%, equal-tailed credible interval for the initial liquid level is $[11.5, 12.4]$\,cm, containing the held-out initial water level of 12\,cm.
The posterior distribution for the drain start time matches its prior distribution, reflecting that the observed final liquid level provides no additional information about the drainage onset time. 
Consequently, uncertainty in the drain start time propagates directly into uncertainty in the inferred initial liquid level. 
Note the negative correlation between the initial liquid level and the drain commence time in the posterior: an earlier drain start implies a longer drainage duration and therefore requires a higher initial liquid level to reach the observed final state, whereas a later drain start requires a lower initial liquid level. 
As a result, bias in the prior distribution for the drain start time would induce corresponding bias in the inferred initial liquid level.

\subsubsection{Two additional problem instances with shorter and longer drainage}
\begin{figure}[!htbp]
        \centering
        	\includegraphics[width=0.8\textwidth]{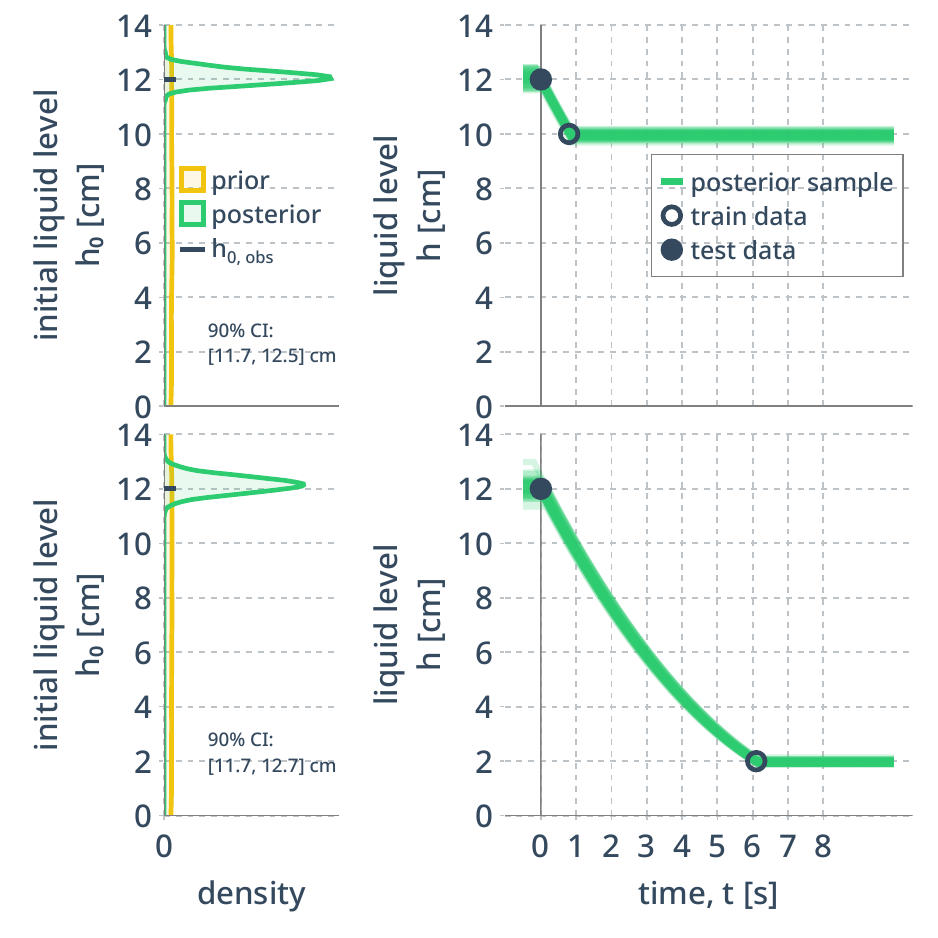}
        \caption{\textbf{Two additional instances of the inverse problem with a shorter (top) and longer (bottom) drainage duration.} Left: Prior and posterior distribution over the initial water level. Right: Water-level trajectories sampled from the posterior compared with the observed final condition and held-out initial condition.
       }\label{fig:ill_cond}
\end{figure}

Next, we tackle two additional instances of the reconstruction problem by considering pollution scenarios with a shorter and longer draining duration. 
Fig.~\ref{fig:ill_cond} shows the solutions to both reconstruction problems. 
The post-drainage observations for the short and long duration problems are $\mathcal{D}^\prime=\{(0.84~\text{min}, 10~\text{cm})\}$ and $\mathcal{D}^\prime=\{(6.1~\text{min}, 2~\text{cm})\}$, respectively. The held-out initial liquid level (12\,cm) falls within the 90\% credible interval of the marginal posteriors in each reconstruction scenario. 
However, uncertainty in the time-reversed water level trajectories increases as the liquid level is reversed backwards further in time. 
The wider 90\% equal-tailed credible interval for the initial liquid level in the longer drainage event, compared to the shorter one, reflects this. 
The effect arises because uncertainty in the observed final liquid level is amplified under backward propagation through the dynamics governed by $h_{\bm \theta}(t)$; it owes to the shape of $h_{\bm \theta}(t)$.
In the limiting case where the tank fully drains, the initial liquid level cannot be uniquely inferred from observation of the final liquid level alone.

\section{Conclusions and Discussion}
This study addressed the inverse problem of inferring the initial liquid level in a partially drained tank from the observed final liquid level and an estimate of the drainage duration. 
To solve this problem, we combined a physics-based drainage model, an empirical model discrepancy function, and experimental liquid level time series data with a Bayesian statistical inversion framework. 
We demonstrated that we can accurately infer the initial liquid level with quantified uncertainty through experiments of a tank draining of water.
Additionally, we showed that uncertainty in the initial liquid level increases with drainage duration and that prior information about the drainage start time is essential, since the observed final liquid level alone cannot identify the drainage onset time.

The proposed methodology has several broader implications and applications.
First, it provides a practical framework for environmental forensics involving partially drained storage tanks when an estimate of the drainage duration is available.
More fundamentally, this work contributes to the study of inverse problems involving draining tanks and related dynamical systems~\cite{groetsch1993inverse_tl,aberman2017dip,fabusola2025inferring}.
Finally, given the accessibility and low cost of the experimental setup, a simplified version of the problem may also be suitable for undergraduate classroom activities illustrating concepts in inverse problems, dynamic modeling, model discrepancy, and Bayesian inference~\cite{groetsch1999inverse,martinez2022notion,savage2021draining}.

Several directions for future work remain. 
First, more sophisticated flow models are of interest for viscous fluids or larger drainage holes where Torricelli’s law may no longer apply~\cite{joye2003tank}. 
Second, relaxing the assumption of atmospheric headspace pressure would require treatment of glugging effects~\cite{mayer2019bottle,clanet2004glug}. 
Third, additional observations from the pollution event, such as the area covered by the discharged liquid or the landing distance of the liquid jet, could provide supplementary information about the initial liquid level~\cite{raja2019efficacy,groetsch1999inverse}.

\section*{Data and code availability}
The data and Julia code to fully reproduce our work are available on Github at \url{github.com/SimonEnsemble/tank_inverse_problem}.

\section*{Acknowledgements}
Thank you Penelope Yong for help with \texttt{Turing.jl} and the developers of open-source software \texttt{CairoMakie.jl}~\cite{danisch2021makie} and \texttt{Turing.jl}~\cite{ge2018turing}. 

\bibliography{refs}

@article{hanford,
author = {Peterson, Reid A. and Buck, Edgar C. and Chun, Jaehun and Daniel, Richard C. and Herting, Daniel L. and Ilton, Eugene S. and Lumetta, Gregg J. and Clark, Sue B.},
title = {Review of the Scientific Understanding of Radioactive Waste at the {U.S. DOE Hanford Site}},
journal = {Environmental Science \& Technology},
volume = {52},
number = {2},
pages = {381-396},
year = {2018}
}

@book{lorentz2012bernstein,
  title={Bernstein polynomials},
  author={Lorentz, George G},
  year={2012},
  publisher={American Mathematical Soc.}
}

@article{betancourt2017conceptual,
  title={A conceptual introduction to {Hamiltonian Monte Carlo}},
  author={Betancourt, Michael},
  journal={arXiv preprint arXiv:1701.02434},
  year={2017}
}

@article{hoffman2014no,
  title={The {No-U-Turn} sampler: adaptively setting path lengths in {Hamiltonian Monte Carlo}},
  author={Hoffman, Matthew D and Gelman, Andrew and others},
  journal={Journal of Machine Learning Research},
  volume={15},
  number={1},
  pages={1593--1623},
  year={2014}
}

@article{rackauckas2017differentialequations,
  title={Differential{E}quations.jl--a performant and feature-rich ecosystem for solving differential equations in {J}ulia},
  author={Rackauckas, Christopher and Nie, Qing},
  journal={Journal of Open Research Software},
  volume={5},
  number={1},
  year={2017},
  publisher={Ubiquity Press}
}

@book{debook,
  title     = "Differential Equations and Linear Algebra",
  author    = "Grant B. Gustafson",
  year      = "2022"
}

@article{aberman2017dip,
  title={Dip transform for {3D} shape reconstruction},
  author={Aberman, Kfir and Katzir, Oren and Zhou, Qiang and Luo, Zegang and Sharf, Andrei and Greif, Chen and Chen, Baoquan and Cohen-Or, Daniel},
  journal={ACM Transactions on Graphics},
  volume={36},
  number={4},
  pages={1--11},
  year={2017},
  publisher={ACM New York, NY, USA}
}

@inproceedings{sidebotham2025workshops,
  title={Workshops for Active Learning and the Draining Tank As a Case Study},
  author={Sidebotham, George and Saissi, Aymane and Wright, Kamau and Feier, Ioan},
  booktitle={ASME International Mechanical Engineering Congress and Exposition},
  volume={89381},
  pages={V007T10A037},
  year={2025},
  organization={American Society of Mechanical Engineers}
}

@article{joye2003tank,
  title={The tank drainage problem revisited: Do these equations actually work?},
  author={Joye, Donald D and Barrett, Branden C},
  journal={The Canadian Journal of Chemical Engineering},
  volume={81},
  number={5},
  pages={1052--1057},
  year={2003},
  publisher={Wiley Online Library}
}

@inproceedings{savage2021draining,
  title={The draining of a tank: a lab experiment in fluid mechanics},
  author={Savage, Dominique and Porterfield, Trent and Penney, W Roy and Clausen, Edgar C},
  booktitle={2021 ASEE Midwest Section Conference},
  year={2021}
}

@book{groetsch1999inverse,
  title={Inverse problems: activities for undergraduates},
  author={Groetsch, Charles W},
  volume={12},
  year={1999},
  publisher={Cambridge University Press}
}

@article{martinez2022notion,
  title={About the notion of inverse problem in {STEM} education},
  author={Martinez-Luaces, Victor and Fern{\'a}ndez-Plaza, Jos{\'e} Antonio and Rico, Luis},
  journal={Active Learning-Research and Practice for STEAM and Social Sciences Education},
  pages={31},
  year={2022},
  publisher={IntechOpen}
}

@article{roy2020convergence,
  title={Convergence diagnostics for {M}arkov chain {M}onte {C}arlo},
  author={Roy, Vivekananda},
  journal={Annual Review of Statistics and Its Application},
  volume={7},
  number={1},
  pages={387--412},
  year={2020},
  publisher={Annual Reviews}
}

@article{clanet2004glug,
  title={On the glug-glug of ideal bottles},
  author={Clanet, Christophe and Searby, Geoffrey},
  journal={Journal of Fluid Mechanics},
  volume={510},
  pages={145--168},
  year={2004},
  publisher={Cambridge University Press}
}

@article{mayer2019bottle,
  title={Bottle emptying: A fluid mechanics and measurements exercise for engineering undergraduate students},
  author={Mayer, Hans C},
  journal={Fluids},
  volume={4},
  number={4},
  pages={183},
  year={2019},
  publisher={MDPI}
}

@article{brooks1998general,
  title={General methods for monitoring convergence of iterative simulations},
  author={Brooks, Stephen P and Gelman, Andrew},
  journal={Journal of Computational and Graphical Statistics},
  volume={7},
  number={4},
  pages={434--455},
  year={1998},
  publisher={Taylor \& Francis}
}

@article{gelman1992inference,
  title={Inference from iterative simulation using multiple sequences},
  author={Gelman, Andrew and Rubin, Donald B},
  journal={Statistical Science},
  volume={7},
  number={4},
  pages={457--472},
  year={1992},
  publisher={Institute of Mathematical Statistics}
}

@article{groetsch1993inverse_tl,
  title={Inverse problems and {T}orricelli's law},
  author={Groetsch, CW},
  journal={The College Mathematics Journal},
  volume={24},
  number={3},
  pages={210--217},
  year={1993},
  publisher={Taylor \& Francis}
}

@Inbook{Calero2008,
title="Discharge from Vessels and Tanks",
bookTitle="The Genesis of Fluid Mechanics, 1640--1780",
year="2008",
publisher="Springer Netherlands",
address="Dordrecht",
pages="271--292",
doi="10.1007/978-1-4020-6414-2_6",
}

@article{schenk2026framework,
  title={A Framework for the Bayesian Calibration of Complex and Data-Scarce Models in Applied Sciences},
  author={Schenk, Christina and Romero, Ignacio},
  journal={arXiv preprint arXiv:2601.22890},
  year={2026}
}

@article{vehtari2021rank,
  title={Rank-normalization, folding, and localization: An improved $\hat{R}$ for assessing convergence of {MCMC} (with discussion)},
  author={Vehtari, Aki and Gelman, Andrew and Simpson, Daniel and Carpenter, Bob and B{\"u}rkner, Paul-Christian},
  journal={Bayesian Analysis},
  volume={16},
  number={2},
  pages={667--718},
  year={2021},
  publisher={International Society for Bayesian Analysis}
}

@article{liu2008drainage,
  title={Drainage and filling in cylindrical and rectangular containers},
  author={Liu, TS and Merati, P and Woodiga, SA and Davis, C and Leong, CH and Johnson, J and Chen, KH},
  journal={Proceedings of the Institution of Mechanical Engineers, Part D: Journal of Automobile Engineering},
  volume={222},
  number={4},
  pages={565--577},
  year={2008},
  publisher={Sage Publications Sage UK: London, England}
}

@article{van2021bayesian,
  title = {Bayesian statistics and modelling},
  volume = {1},
  number = {1},
  journal = {Nature Reviews Methods Primers},
  author = {van de Schoot,  Rens and Depaoli,  Sarah and King,  Ruth and Kramer,  Bianca and M\"{a}rtens,  Kaspar and Tadesse,  Mahlet G. and Vannucci,  Marina and Gelman,  Andrew and Veen,  Duco and Willemsen,  Joukje and Yau,  Christopher},
  year = {2021},
  month = {January} 
}

@incollection{dashti2015bayesian,
  title={The {B}ayesian approach to inverse problems},
  author={Dashti, Masoumeh and Stuart, Andrew M},
  booktitle={Handbook of Uncertainty Quantification},
  pages={1--118},
  year={2015},
  publisher={Springer}
}

@article{mebane2013bayesian,
  title={Bayesian calibration of thermodynamic models for the uptake of {CO}$_2$ in supported amine sorbents using ab initio priors},
  author={Mebane, David S and Bhat, K Sham and Kress, Joel D and Fauth, Daniel J and Gray, McMahan L and Lee, Andrew and Miller, David C},
  journal={Physical Chemistry Chemical Physics},
  volume={15},
  number={12},
  pages={4355--4366},
  year={2013},
  publisher={Royal Society of Chemistry}
}

@article{johnston2014updating,
  title={Updating exposure models of indoor air pollution due to vapor intrusion: {B}ayesian calibration of the {Johnson-Ettinger} model},
  author={Johnston, Jill E and Sun, Qiang and Gibson, Jacqueline MacDonald},
  journal={Environmental Science \& Technology},
  volume={48},
  number={4},
  pages={2130--2138},
  year={2014},
  publisher={ACS Publications}
}

@article{hou2021review,
  title={Review on building energy model calibration by {B}ayesian inference},
  author={Hou, Danlin and Hassan, IG and Wang, L},
  journal={Renewable \& Sustainable Energy Reviews},
  volume={143},
  pages={110930},
  year={2021},
  publisher={Elsevier}
}

@article{mebane2023bayesian,
  title={Bayesian calibration of electrical conductivity relaxation and isotope exchange-secondary ion mass spectrometry experiments},
  author={Mebane, David S},
  journal={Journal of Electroceramics},
  volume={51},
  number={4},
  pages={239--245},
  year={2023},
  publisher={Springer}
}

@article{calvetti2018inverse,
  title={Inverse problems: From regularization to {B}ayesian inference},
  author={Calvetti, Daniela and Somersalo, Erkki},
  journal={Wiley Interdisciplinary Reviews: Computational Statistics},
  volume={10},
  number={3},
  pages={e1427},
  year={2018},
  publisher={Wiley Online Library}
}

@article{kennedy2001bayesian,
  title={Bayesian calibration of computer models},
  author={Kennedy, Marc C and O'Hagan, Anthony},
  journal={Journal of the Royal Statistical Society: Series B (Statistical Methodology)},
  volume={63},
  number={3},
  pages={425--464},
  year={2001},
  publisher={Wiley Online Library}
}

@article{brynjarsdottir2014learning,
  title={Learning about physical parameters: The importance of model discrepancy},
  author={Brynjarsdottir, Jenny and O'Hagan, Anthony},
  journal={Inverse Problems},
  volume={30},
  number={11},
  pages={114007},
  year={2014},
  publisher={IOP Publishing}
}

@book{kaipio2006statistical,
  title={Statistical and Computational Inverse Problems},
  author={Kaipio, Jari and Somersalo, Erkki},
  volume={160},
  year={2006},
  publisher={Springer Science \& Business Media}
}

@article{waqar2023tutorial,
  title={A tutorial on the {B}ayesian statistical approach to inverse problems},
  author={Waqar, Faaiq G and Patel, Swati and Simon, Cory M},
  journal={APL Machine Learning},
  volume={1},
  number={4},
  year={2023},
  publisher={AIP Publishing}
}

@article{fabusola2025inferring,
  title={Inferring the cross-sectional area profile of an unseen solid in a draining tank from liquid level dynamics},
  author={Fabusola, Gbenga and Simon, Cory M},
  journal={Chemical Engineering Science},
  volume={309},
  pages={121488},
  year={2025},
  publisher={Elsevier}
}

@article{cullinan2002epidemiological,
  title={Epidemiological assessment of health effects from chemical incidents},
  author={Cullinan, Paul},
  journal={Occupational and Environmental Medicine},
  volume={59},
  number={8},
  pages={568--572},
  year={2002},
  publisher={BMJ Publishing Group Ltd}
}

@article{guerin2014understanding,
  title={Understanding causes of leaking plant and equipment on construction sites that can lead to soil and groundwater contamination},
  author={Guerin, Turlough},
  journal={Remediation Journal},
  volume={25},
  number={1},
  pages={115--131},
  year={2014},
  publisher={Wiley Online Library}
}

@book{pullarcot2015above,
  title={Above Ground Storage Tanks},
  author={Pullarcot, Sunil},
  year={2015},
  publisher={CRC Press}
}

@article{doi:10.1021/acsomega.3c05187,
author = {Zhang, Hao and Yang, Yang and Ma, Shaobing and Yuan, Wenchao and Gao, Mingjun and Li, Tongtong and Wei, Yuquan and Wang, Yanwei and Xiong, Yanna and Li, Aiyang and Zhao, Bin},
title = {Development of a Multifaceted Perspective for Systematic Analysis, Assessment, and Performance for Environmental Standards of Contaminated Sites},
journal = {ACS Omega},
volume = {9},
number = {3},
pages = {3078-3091},
year = {2024},
}

@article{williams2021vessel,
  title={Vessel drainage under the influence of gravity},
  author={Williams, Hollis},
  journal={The Physics Teacher},
  volume={59},
  number={8},
  pages={629--631},
  year={2021},
  publisher={AIP Publishing}
}

@article{pavesi2019investigating,
  title={Investigating {T}orricelli’s law (and more) with a 19th-century bottle},
  author={Pavesi, Laura},
  journal={The Physics Teacher},
  volume={57},
  number={2},
  pages={106--108},
  year={2019},
  publisher={AIP Publishing}
}

@article{belinskiy2020time,
  title={Time Optimization of a Draining Tank and Some Similar Problems on Star Graphs},
  author={Belinskiy, Boris P and White, Douglas C},
  journal={Punjab University Journal of Mathematics},
  volume={51},
  number={7},
  year={2020}
}

@article{rother2024modelling,
  title={Modelling tank drainage using a simple apparatus},
  author={Rother, Michael A},
  journal={International Journal of Mathematical Education in Science and Technology},
  volume={55},
  number={2},
  pages={295--307},
  year={2024},
  publisher={Taylor \& Francis}
}

@article{ly2021site,
  title={On-Site Detection for Hazardous Materials in Chemical Accidents},
  author={Ly, Nguyen Ho{\`a}ng and Kim, Ho Hyun and Joo, Sang-Woo},
  journal={Bulletin of the Korean Chemical Society},
  volume={42},
  number={1},
  pages={4--16},
  year={2021},
  publisher={Wiley Online Library}
}

@article{danisch2021makie,
  title={Makie. jl: Flexible high-performance data visualization for {J}ulia},
  author={Danisch, Simon and Krumbiegel, Julius},
  journal={Journal of Open Source Software},
  volume={6},
  number={65},
  pages={3349},
  year={2021}
}

@article{bezanson2017julia,
  title={Julia: A fresh approach to numerical computing},
  author={Bezanson, Jeff and Edelman, Alan and Karpinski, Stefan and Shah, Viral B},
  journal={SIAM Review},
  volume={59},
  number={1},
  pages={65--98},
  year={2017},
  publisher={SIAM}
}

@article{ramirez1996detection,
  title={Detection of leaks in underground storage tanks using electrical resistance methods},
  author={Ramirez, Abelardo and Daily, William and Binley, Andrew and LaBrecque, Douglas and Roelant, David},
  journal={Journal of Environmental and Engineering Geophysics},
  volume={1},
  number={3},
  pages={189--203},
  year={1996},
  publisher={Society of Exploration Geophysicists}
}

@techreport{ho2001review,
  title={Review of chemical sensors for in-situ monitoring of volatile contaminants},
  author={Ho, Clifford K and Itamura, Michael T and Kelley, Michael J and Hughes, Robert C},
  year={2001},
  institution={Sandia National Lab. Albuquerque, NM (United States)}
}

@inproceedings{ge2018turing,
  title={Turing: a language for flexible probabilistic inference},
  author={Ge, Hong and Xu, Kai and Ghahramani, Zoubin},
  booktitle={International conference on artificial intelligence and statistics},
  pages={1682--1690},
  year={2018},
  organization={PMLR}
}

@article{mclaughlin2016spills,
  title={Spills of hydraulic fracturing chemicals on agricultural topsoil: biodegradation, sorption, and co-contaminant interactions},
  author={McLaughlin, Molly C and Borch, Thomas and Blotevogel, Jens},
  journal={Environmental Science \& Technology},
  volume={50},
  number={11},
  pages={6071--6078},
  year={2016},
  publisher={ACS Publications}
}

@article{bongers2008challenges,
  title={Challenges of exposure assessment for health studies in the aftermath of chemical incidents and disasters},
  author={Bongers, Sim and Janssen, Nicole AH and Reiss, B and Grievink, Linda and Lebret, Erik and Kromhout, Hans},
  journal={Journal of Exposure Science \& Environmental Epidemiology},
  volume={18},
  number={4},
  pages={341--359},
  year={2008},
  publisher={Nature Publishing Group}
}

@inproceedings{hongguang2014fault,
  title={Fault tree analysis of the storage tanks in the chemical industry},
  author={Hongguang, Ao},
  booktitle={2014 Fourth International Conference on Instrumentation and Measurement, Computer, Communication and Control},
  pages={928--931},
  year={2014},
  organization={IEEE}
}

@article{powell2012carrying,
  title={Carrying biomath education in a leaky bucket},
  author={Powell, James A and Kohler, Brynja R and Haefner, James W and Bodily, Janice},
  journal={Bulletin of Mathematical Biology},
  volume={74},
  number={9},
  pages={2232--2264},
  year={2012},
  publisher={Springer}
}

@article{montalvo2021environmental,
  title={Environmental impact of wine fermentation in steel and concrete tanks},
  author={Montalvo, Francisco Flor and Garc{\'\i}a-Alcaraz, Jorge Luis and C{\'a}mara, Eduardo Mart{\'\i}nez and Jim{\'e}nez-Mac{\'\i}as, Emilio and Blanco-Fern{\'a}ndez, Julio},
  journal={Journal of Cleaner Production},
  volume={278},
  pages={123602},
  year={2021},
  publisher={Elsevier}
}

@article{chang2006study,
  title={A study of storage tank accidents},
  author={Chang, James I and Lin, Cheng-Chung},
  journal={Journal of Loss Prevention in the Process Industries},
  volume={19},
  number={1},
  pages={51--59},
  year={2006},
  publisher={Elsevier}
}

@article{raja2019efficacy,
  title={Efficacy of existing transient models for spill area forecasting},
  author={Raja, S and Reddy, TLP and Tauseef, SM and SA, Abbasi and others},
  journal={Journal of Chemical Health \& Safety},
  volume={26},
  number={4-5},
  pages={33--37},
  year={2019},
  publisher={Elsevier}
}

@article{travnivcek2019prevention,
  title={Prevention of accidents to storage tanks for liquid products used in agriculture},
  author={Tr{\'a}vn{\'\i}{\v{c}}ek, Petr and Kotek, Lubo{\v{s}} and Junga, Petr and Koutn{\`y}, Tom{\'a}{\v{s}} and Novotn{\'a}, Jana and V{\'\i}t{\v{e}}z, Tom{\'a}{\v{s}}},
  journal={Process Safety \& Environmental Protection},
  volume={128},
  pages={193--202},
  year={2019},
  publisher={Elsevier}
}

@misc{iea,
  title = {{Fukushima Daiichi ALPS Treated Water Discharge - FAQs}},
  author={International Atomic Energy Agency},
  note = {Accessed: 2025-04-19}
}

@article{huang2013technical,
  title={Technical aspects of storage tank loss prevention},
  author={Huang, Szu-Ying and Mannan, M Sam},
  journal={Process Safety Progress},
  volume={32},
  number={1},
  pages={28--36},
  year={2013},
  publisher={Wiley Online Library}
}

@article{sylvester2013radioactive,
  title={Radioactive liquid waste treatment at {Fukushima Daiichi}},
  author={Sylvester, Paul and Milner, Tim and Jensen, Jesse},
  journal={Journal of Chemical Technology \& Biotechnology},
  volume={88},
  number={9},
  pages={1592--1596},
  year={2013},
  publisher={Wiley Online Library}
}

@misc{liu2014bayes,
  author       = {Liu, Han and Wasserman, Larry},
  title        = {Statistical Machine Learning: Bayesian Inference},
  year         = {2014},
  note         = {Lecture notes, Carnegie Mellon University},
  url          = {https://www.stat.cmu.edu/~larry/=sml/Bayes.pdf},
  urldate      = {2026-02-26}
}
\bibliographystyle{unsrt}
\clearpage
\appendix

\renewcommand{\thefigure}{S\arabic{figure}}
\renewcommand{\thetable}{S\arabic{table}}
\renewcommand{\theequation}{S\arabic{equation}}

\setcounter{figure}{0}
\setcounter{table}{0}
\setcounter{equation}{0}

\section*{Supplementary Information}
\addcontentsline{toc}{section}{Supplementary Information}
\subsection*{Bernstein basis polynomials} 

\begin{figure}[H]
    \centering
      \includegraphics[width=0.65\textwidth]{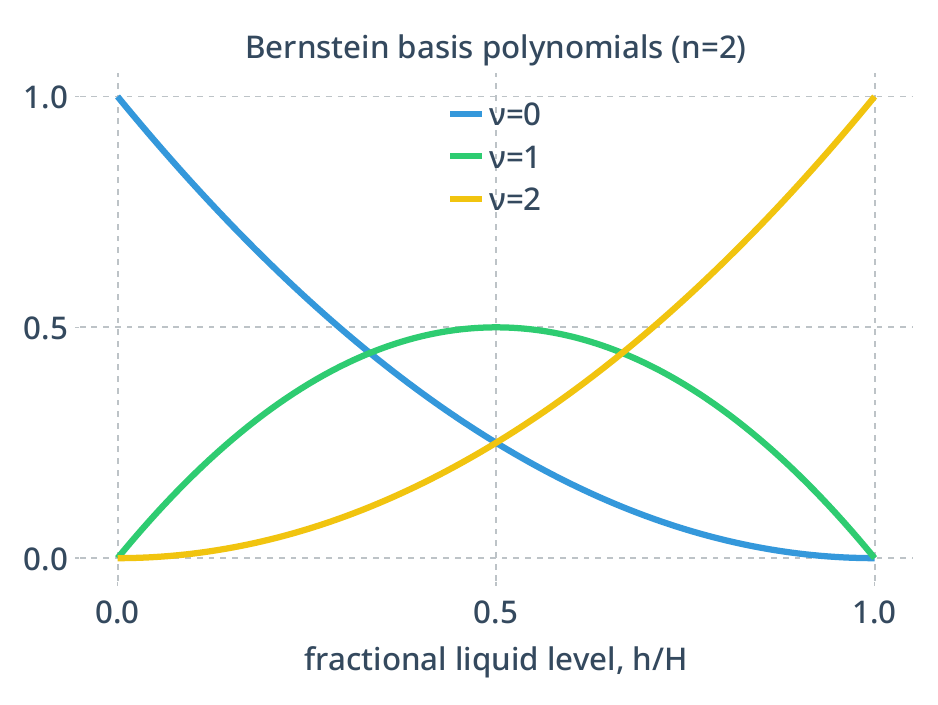}
    \caption{
    \textbf{Bernstein basis polynomials} for $n=2$.
    } \label{fig:bern}
\end{figure}

\clearpage

\subsection*{Prior distributions} \label{sec:detailed_priors}
\paragraph{Tank geometry and orifice.}
The tank's height, $h_{\max}$, was measured to be approximately 14 cm using a measuring tape.
The top side length $x_t$ was measured as 8.7 cm, and the bottom side length $x_b$ as 8.4 cm.
The radius of the orifice, $r$, was estimated to be 0.12 cm based on the drill bit used to create the hole.
\begin{subequations} \label{eq:geom_priors}
	\begin{align}
    		H_{\max} &\sim \mathcal{N}\left(14.0~\text{cm}, (0.1~\text{cm})^2\right), \label{eq:H_max_prior} \\
    		X_t &\sim \mathcal{N}\left(8.7~\text{cm}, (0.1~\text{cm})^2\right), \label{eq:Xt_prior} \\
    		X_b &\sim \mathcal{N}\left(8.4~\text{cm}, (0.1~\text{cm})^2\right), \label{eq:Xb_prior} \\
    		r &\sim \mathcal{N}\left(0.12~\text{cm}, \left[(0.05)(0.12~\text{cm})\right]^2\right). \label{eq:R0_prior}
	\end{align}
\end{subequations}
The priors in eqns.~\eqref{eq:H_max_prior}-\eqref{eq:Xb_prior} are centered on the measured values, with a standard deviation of 0.1 cm to account for uncertainty arising from the spacing between consecutive markings on the measuring tape.
The prior in eqn.~\eqref{eq:R0_prior} for the orifice radius incorporates a 5\% relative uncertainty.

\paragraph{Discharge coefficient.}
The discharge coefficient $C$ is a dimensionless quantity defined on the open interval $[0,1]$.
Accordingly, a Beta distribution is an appropriate prior. 
For low-viscosity liquids flowing through a circular orifice, the expected value of $C$ is approximately 0.6, which motivates the following prior specification:
\begin{equation*}
	C \sim \mathrm{Beta}(6,4).
\end{equation*}
This parameterization yields a prior mean of 0.6 and a standard deviation of approximately 0.15. 
The relatively diffuse prior enables the liquid-level time-series data to inform the posterior distribution of $C$ without imposing excessive constraints.

\paragraph{Measurement noise.}
Since $\sigma$ is strictly positive by definition, an exponential prior is imposed.
While any prior distribution with support on the nonnegative real line, such as the Gamma or log-normal, could be used for $\sigma$, the exponential distribution is chosen for its simplicity and its capacity to encode weak prior information.
The following prior is therefore adopted:
\begin{equation*}
	\Sigma \sim \mathrm{Exponential}\left(4\text{ cm}^{-1}\right). 
\end{equation*}
The scale of this prior is selected to correspond to the measurement resolution.
The relatively large standard deviation indicates substantial uncertainty in this value, allowing the data to exert a dominant influence on the posterior distribution.

\paragraph{Model discrepancy.}
Zero-centered Laplace priors are assigned to the model discrepancy parameters in $\delta_{\bf a}(\cdot)$, reflecting the assumption that the physics-based model is roughly accurate. 
These priors favor small discrepancies unless data indicate otherwise, capturing this model assumption. 
The Laplace distribution promotes sparsity—favoring minor corrections but allowing larger ones if justified by evidence.
The parameterization of these priors are:
\begin{equation}
	A_i \sim \text{Laplace}(0~\text{cm}, 0.25~\text{cm}) \text{ for } i \in \{0, 1,2\}. \nonumber
\end{equation}

\paragraph{Initial conditions.}
For the initial conditions of the three tank-draining experiments, we treat the pollution scenario distinctly from the calibration experiments.

\subparagraph{Pollution scenario.}
For the pollution scenario, a diffuse prior is placed over the initial liquid level using a uniform distribution, reflecting no prior knowledge about the amount of liquid in the tank before draining began:
\begin{equation*}
	H_0\mid H_{\max} \sim \mathcal{U}(0, H_{\max}).
\end{equation*}
This approach allows for any initial liquid volume between empty and full, without bias toward a specific value. 
Because $H_{\max}$ is uncertain, the prior is conditioned on this parameter. 
The observed final liquid level subsequently informs the inference process.

An informative prior distribution is specified for the drain start time:
\begin{equation*}
	T_0 \sim \mathcal{N}(0~\text{s}, (5~\text{s})^2).
\end{equation*}
Based on the estimated drain start time, a $\pm$5 s uncertainty is assumed around $t_0 =$ 0 s. 
This prior may be updated if additional information becomes available, such as ``the tank was at least half full" or ``it drained between 200 and 300 s." 

\subparagraph{Calibration experiments.}
For the calibration experiments, we impose informative priors on the drain start time and the initial liquid level, since these quantities were controlled in the experimental setup.
The priors for the initial conditions are:
\begin{subequations} \label{eq:ic_priors}
	\begin{align}
  		T_0^{(1)},\, T_0^{(2)} &\stackrel{\rm i.i.d.}{\sim} \mathcal{N}(0, (0.25~\text{s})^2), \nonumber \\
  		H_0^{(1)} \mid \sigma,\, H_0^{(2)} \mid \sigma &\stackrel{\rm i.i.d.}{\sim} \mathcal{N}(14.0~\text{cm}, \sigma^2), \nonumber \label{eq:H0_prior}
	\end{align}
\end{subequations}
where i.i.d. denotes independently and identically distributed.
Thus, we admit uncertainty in the initial conditions imposed at the start of each experiment and allow for the possibility that the two experiments did not begin from exactly identical states.

\subsection*{Posterior convergence diagnostics}

\begin{figure}[H]
    \centering
      \includegraphics[width=\textwidth]{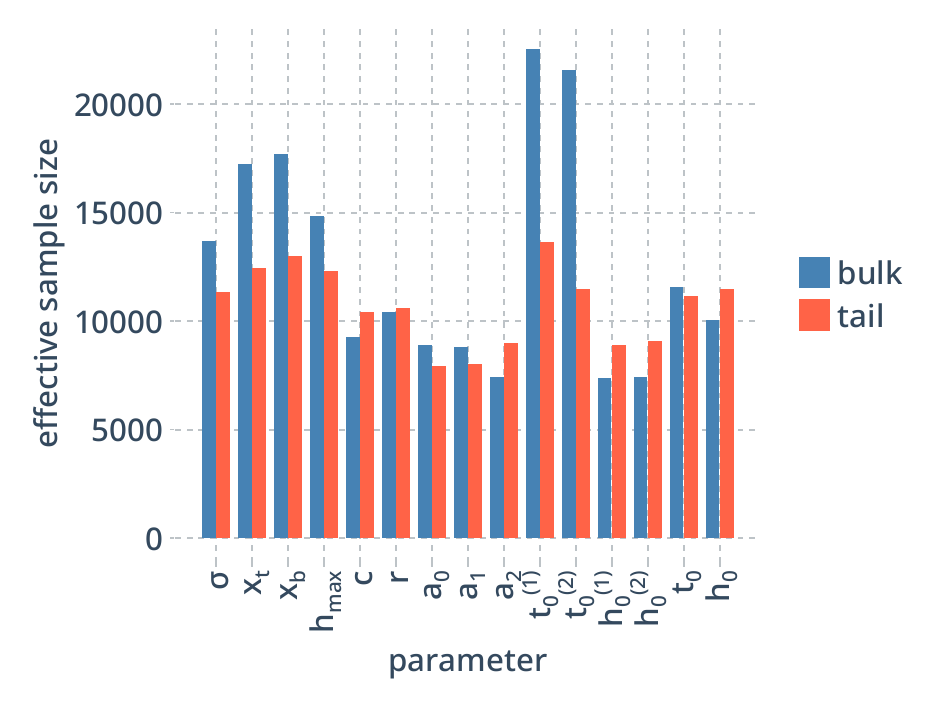}
    \caption{
    \textbf{Effective sample size for the posterior MCMC sampling.}
    } \label{fig:ess}
\end{figure}

\begin{figure}[H]
    \centering
      \includegraphics[width=\textwidth]{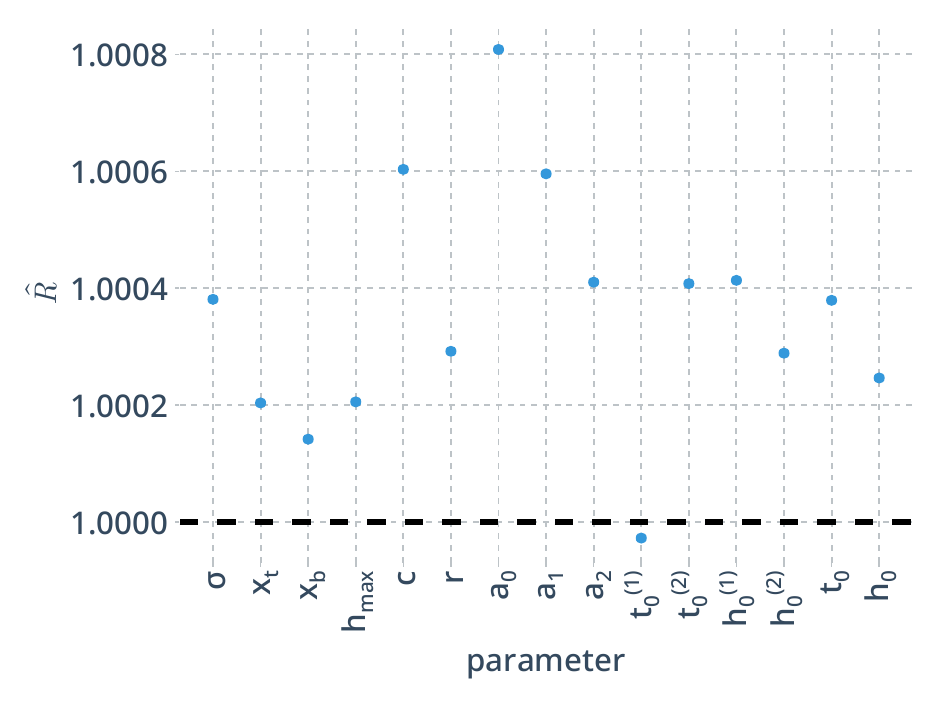}
    \caption{
    \textbf{The $\hat{R}$ statistic for convergence assessment \cite{vehtari2021rank}.}
    } \label{fig:rhat}
\end{figure}

\begin{figure}[H]
    \centering
      \includegraphics[width=0.5\textwidth]{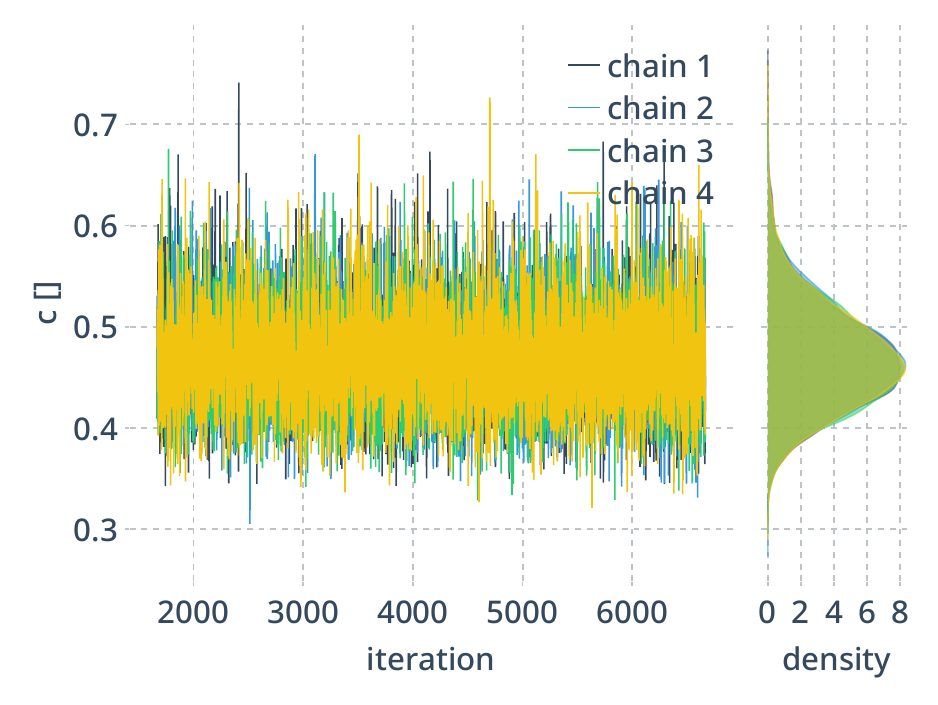}
      \includegraphics[width=0.5\textwidth]{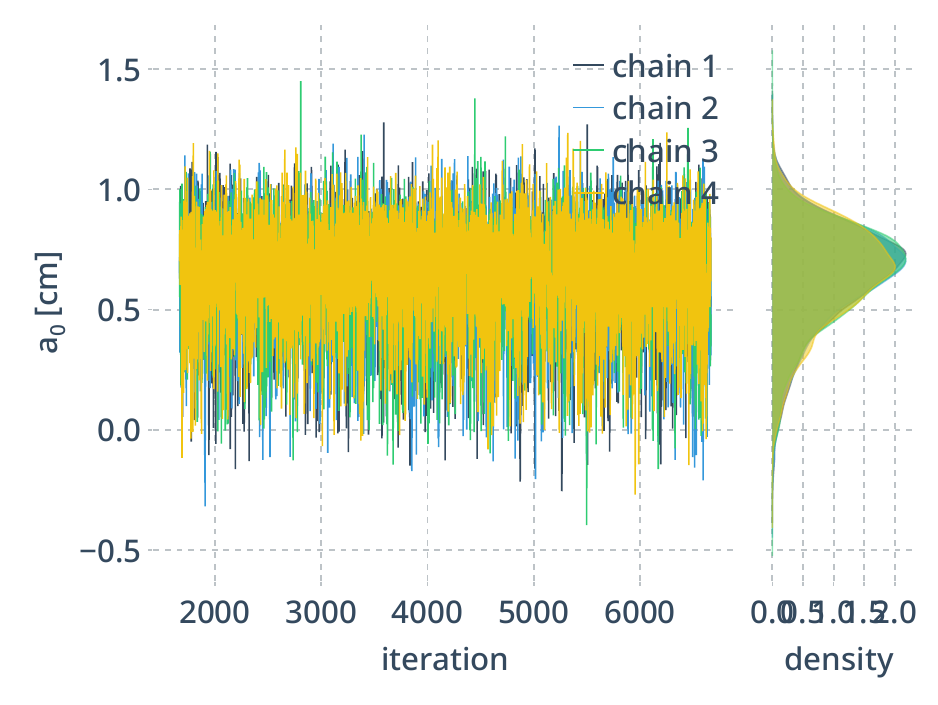}
      \includegraphics[width=0.5\textwidth]{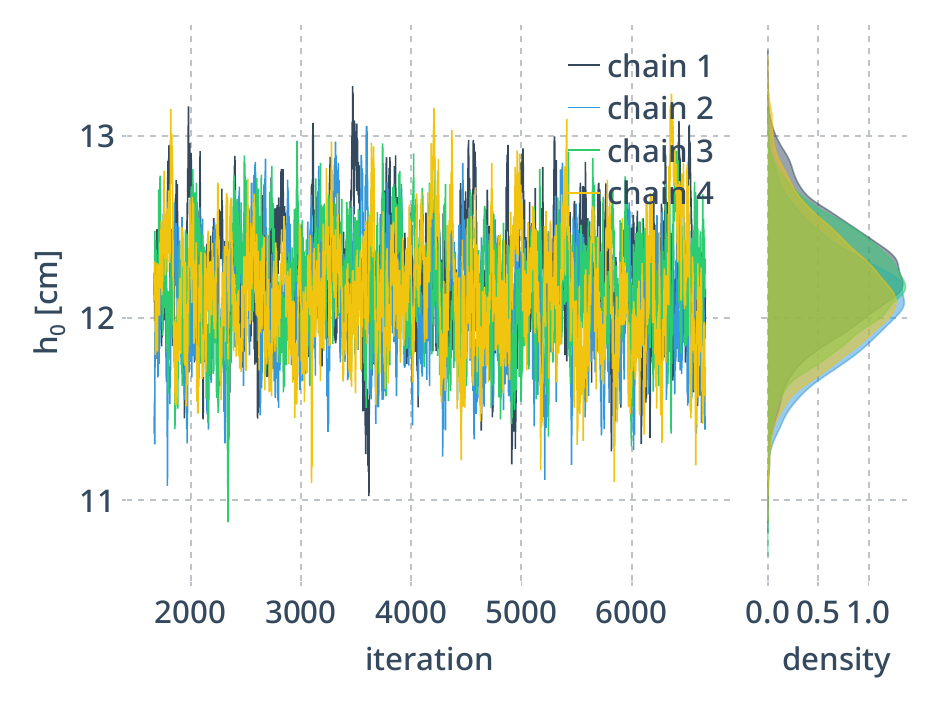}
    \caption{
    \textbf{Example trace plots.}
    For four Markov chains simulated in parallel, we show the trace plot and posterior distribution for a few unknowns. 
    } \label{fig:convergence}
\end{figure}

\subsection*{Posterior correlations}

\begin{figure}[H]
    \centering
      \includegraphics[width=\textwidth]{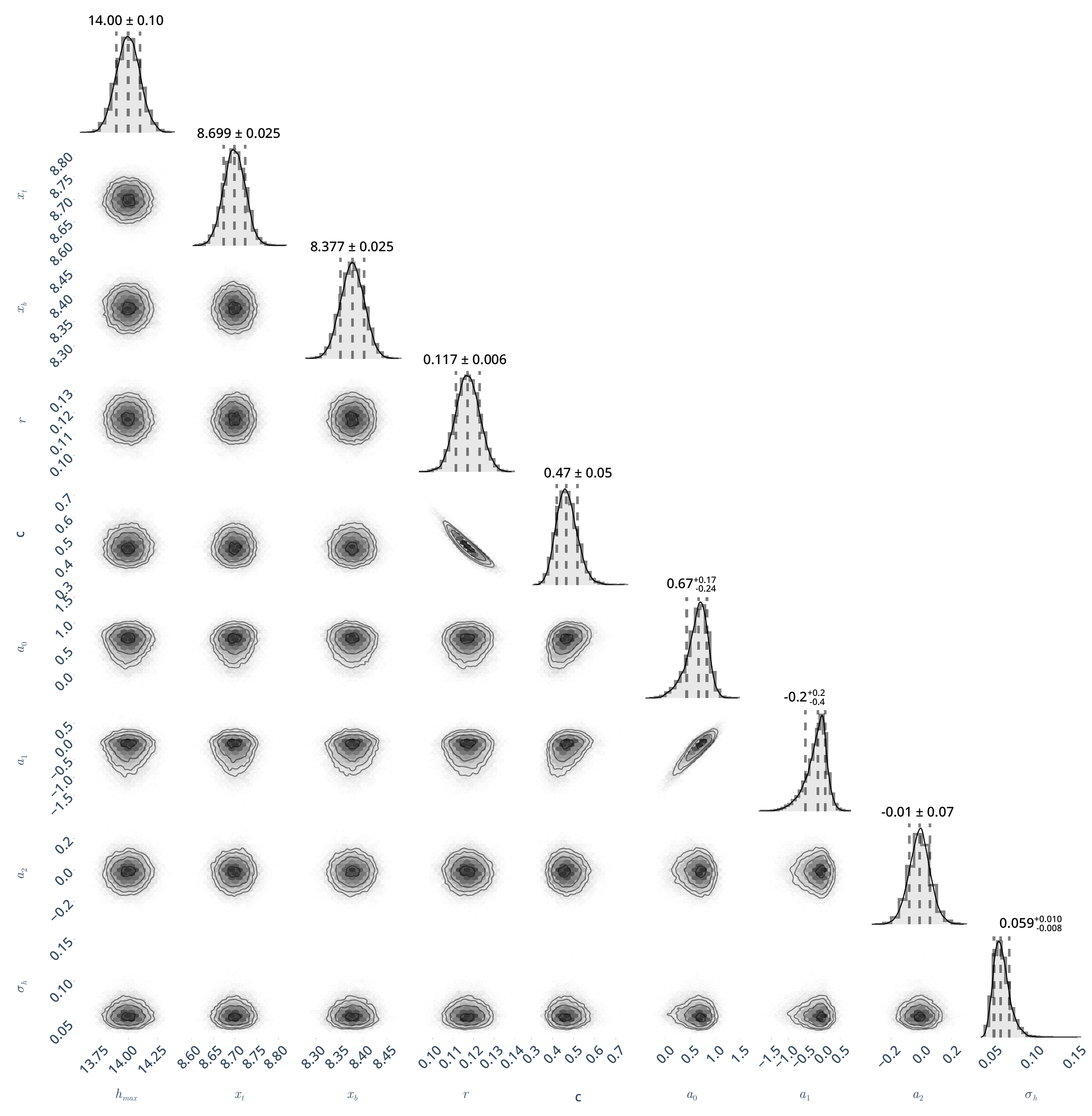}
    \caption{
    \textbf{Pairplot showing the posterior distribution from model calibration in detail. Diagonal: marginal distribution; Off-diagonal: joint distribution.}
    } \label{fig:pairplot_calibration}
\end{figure}


\clearpage

\end{document}